%% file: paper.tex
\documentclass[usenatbib,letterpaper]{mn2e}
\usepackage[totalwidth=480pt, totalheight=680pt]{geometry}
\usepackage{amssymb}
\usepackage{amsmath}
\usepackage{natbib}
\usepackage{graphicx}
\usepackage{color}
\usepackage{rotating}
\usepackage{url}
\usepackage{ifthen}

\newcommand{\Mpch}{{$h^{-1}$~Mpc}}
\newcommand{\kms}{km s$^{-1}$}
\newcommand{\kmsMpc}{{km s$^{-1}$ Mpc$^{-1}$}}
\newcommand{\ktmpp}{K_\text{2M++}}
\newcommand{\LCDM}{$\Lambda$CDM}
\newcommand{\hMsun}{$h^{-1} \mathrm{M}_\odot$}

\newcommand{\ten}[1]{\times 10^{#1}}

\def\proptosima{$\; \buildrel \propto \over \sim \;$}
\def\simpropto{\lower.5ex\hbox{\proptosima}}

\begin{document}

\bibliographystyle{mn2e}

\title{The 2M++ galaxy redshift catalogue }

\author[G. Lavaux \& M.~J. Hudson]{Guilhem Lavaux$^{1,2}$ \& Michael~J. Hudson$^{3,4,5}$\\
  $^{1}$ Department of Physics, University of Illinois at Urbana-Champaign, 1110 West Green Street, Urbana, IL 61801-3080\\
  $^{2}$ Department of Physics and Astronomy, The Johns Hopkins University, 3701 San Martin Drive, Baltimore, MD 21218, USA\\
  $^{3}$ Institut d'Astrophysique de Paris, UMR7095 CNRS, Univ. Pierre et Marie Curie, 98 bis Boulevard Arago, 75014 Paris, France\\
  $^{4}$ Department of Physics \& Astronomy, University of Waterloo, Waterloo, ON,  N2L 3G1 Canada\\
  $^{5}$ Perimeter Institute for Theoretical Physics, 31 Caroline St. N., Waterloo, ON N2L 2Y5, Canada%
}

\date{\today}

\maketitle

\begin{abstract}
  Peculiar velocities arise from gravitational instability, and thus are linked to the surrounding  distribution of matter.
  In order to understand the motion of the Local Group with respect to the Cosmic Microwave Background,
  a deep all-sky map of the galaxy distribution is required.
  Here we present a new redshift compilation of 69~160 galaxies, dubbed 2M++, to
  map large-scale structures of the Local Universe over 
  nearly the whole sky, and reaching depths of $K\le 12.5$, or 200\Mpch.
  The target catalogue is based on the Two-Micron-All-Sky Extended Source Catalog
  (2MASS-XSC). The primary sources of redshifts are the 2MASS
  Redshift Survey, the 6dF galaxy redshift survey and the Sloan
  Digital Sky Survey (DR7).  We assess redshift completeness in each region and 
  compute the weights required to correct for redshift incompleteness and
  apparent magnitude limits, and discuss corrections for incompleteness in the Zone of Avoidance.
  We present the density field for this survey, and discuss the importance of large-scale structures such as the Shapley Concentration. 
\end{abstract}


\section{Introduction}

Peculiar velocities remain the only method to map the distribution of dark matter on very large scales in the low redshift Universe.  Recently, several intriguing measurements \citep{KABKE1, WFH09, Lavaux10, KABKE3, FelWatHud10} of the mean or ``bulk'' flow on scales larger than 100\Mpch{} suggest a high velocity of our local $\sim 100$ \Mpch{} volume with respect to the Cosmic Microwave Background (CMB) frame. In the standard cosmological framework, peculiar velocities are proportional to peculiar acceleration and so one expects the bulk flow to arise from fluctuations in the distribution of matter, and hence presumably of galaxies, on very large scales. 
Another  statistic for measuring such large-scale fluctuations is the convergence of the gravity dipole as a function of distance. However, the rate of convergence has been a subject of recent debate \citep[][and references therein]{KE06,EH06,EL06,Lavaux10,BCMJ11}. A closely-related topic is the gravitational influence of the Shapley Concentration, the largest concentration of galaxy clusters in the nearby Universe.  It is therefore important to have catalogues that are as full sky and as deep as possible to understand whether the distribution of matter in the nearby Universe may explain the above-mentioned results.

It is already possible with currently available data to build a redshift catalogue significantly deeper than previous full-sky galaxy redshift catalogues like PSCz \citep{PSCz} or the Two-Micron-All-Sky Redshift Survey \citep[2MRS]{2MRS,EH06,2MRS115}. We present here a new catalogue called the 2M++ galaxy redshift compilation. The photometry for this compilation is  based is on the Two-Micron-All-Sky-Survey (2MASS) Extended Source Catalog \citep[2MASS-XSC]{2MASS}. We gather the high-quality redshifts from the 2MASS redshift survey \citep{2MRS,EH06,2MRS115} limited to $K=11.5$, the 6dF galaxy redshift survey Data Release Three \citep[6dFGRS][]{6dfDR3} and the Sloan Digital Sky Survey Data Release Seven \citep[SDSS-DR7][]{SDSSDR7}.

A summary of this paper is as follows. In Section~\ref{sec:building}, we describe the steps in constructing
the 2M++ redshift galaxy catalogue: source selection, magnitude
corrections, redshift incompleteness estimation and correction, the
luminosity function (LF) estimation and the final weight computation. In Section~\ref{sec:zoa}, we discuss the Zone of Avoidance (ZoA) in our catalogue, and how its effects can be mitigated. 
In Section~\ref{sec:groups}, we define groups of galaxies and check some of their overall properties.  In
Section~\ref{sec:density}, we compute and analyse the density field, presenting maps of 
the Supergalactic plane and three cluster density and velocity profiles. Section~\ref{sec:conclusion} summarizes our key results.


\section{Catalogue construction}
\label{sec:building}

In this Section, we describe the construction of the 2M++ galaxy
redshift catalogue from the different data sources. First, in
Section~\ref{sec:datasource}, we describe the source data sets that form 
the basis of our catalogue, as well as the primary steps in the construction of the 2M++ catalogue. 
We then present the methodology used
for merging these different sources. In Section~\ref{sec:magcor}, we
describe the corrections applied to apparent magnitudes to homogenize
the target selection. In Section~\ref{sec:red_cloning}, we test and
apply the redshift cloning procedure to our data to increase the
overall redshift completeness. We then estimate redshift completeness
(Section~\ref{sec:red_completeness}) and present the number counts of
galaxies as a function of redshift (Section~\ref{sec:red_number}). Finally, we compute the LF of our sample in Section~\ref{sec:lumfun} and compute the total weights to apply to each galaxy in Section~\ref{sec:weights}.


\subsection{Source datasets and construction procedure}
\label{sec:datasource}

Our catalogue is based on the Two-Micron-All-Sky-Survey \citep[2MASS]{2MASS} photometric catalogue for target selection, which has very high completeness up to $K_S=13.2$ \citep{Cole01}. Hereafter, for brevity,  we use $K$ in place of $K_S$. 
As noted above, we will be using redshifts from the SDSS-DR7, the 6dfGRS and the 2MRS. In addition to these main sources, we have gathered additional redshifts from a number of other sources \citep{red0,red1,red2,red3,red4,red5,red6,red7} through NED queries.\footnote{The NASA/IPAC Extragalactic Database (NED) is operated by the Jet Propulsion Laboratory, California Institute of Technology, under contract with the National Aeronautics and Space Administration.} Due to the
inhomogeneity of the target selection between the different redshift
surveys, we think that it is more appropriate to define a new target
selection rather than using existing target databases from the above
surveys. We used of the NYU-VAGC \citep{NYUVAGC} catalogue for
matching the SDSS data to the 2MASS Extended Source Catalog
(2MASS-XSC). The NYU-VAGC provides the SDSS survey mask in MANGLE format
\citep{mangle}.\footnote{We use the file named \texttt{lss\_combmask.dr72.ply}, which gives the geometry of the DR72 sample in terms of target selection with bright stars excised.} We sampled the mask on a {\sc HEALPix} grid at
$N_\text{side}=512$ ($\sim$10 arcminutes resolution). This angular
resolution corresponds to $\sim$1\Mpch{} at $\sim$300\Mpch{}. Because
ultimately we will be smoothing the density field on scales of $\sim$4\Mpch{}, the mask has sufficient resolution for our purposes. Additionally, we filter out from our target selection the extended sources that are known not to be galaxies.\footnote{We require that the {\sc visual\_{}code} is not equal to two.}

We aim to limit 2M++ at $K \simeq 12.5$ regions of the sky covered by SDSS
or by 6dF. The exact cut depends on the adopted definition for the
magnitude. As we want to retain as much as possible information from
the shallower 2MRS catalogue, we opt to follow closely the magnitude used by 2MRS
for target selection. We define as $\ktmpp$ the magnitude of a
galaxy measured in the $K_\text{S}$ band, within the circular isophote
at 20~mag\,arcsec$^{-2}$, after various corrections as described below (Section~\ref{sec:magcor}).
 Several of the steps taken to build the catalogue are described in greater detail in the
following Sections. We now outline these steps:
\begin{enumerate}
\item We import the redshift information for 2MASS-XSC galaxies from the
  NYU-VAGC for SDSS-DR7, the 6dF-DR3, and from the 2MASS Redshift Survey.
\item We correct for small-scale redshift incompleteness (arising from
  fibre collisions) by `cloning' the redshifts of nearby galaxies
  (Section~\ref{sec:red_cloning}).
\item We correct the apparent magnitudes for Galactic dust extinction
  (Section~\ref{sec:magcor}).
\item We use the redshift to correct for galaxy evolutionary effects and aperture corrections
  (Section~\ref{sec:magcor}). We call this magnitude $\ktmpp$. At those
  magnitudes, we assume that the photometric completeness is one at
  Galactic latitudes higher than 10$^\circ$.
\item We compute two sets of galaxy samples: a target sample
  with $\ktmpp \le 11.5$ in regions not covered by 6dFGRS or SDSS, and a target sample limited to 
  $\ktmpp \le 12.5$ in regions covered by SDSS or 6dFGRS.
\item We estimate the redshift completeness as a function of position on the sky for in each of these regions.
\item We place galaxies in groups and clusters using  a percolation algorithm.
\item We compute the Schechter parameters of the LF
    of the combined catalogue (Section~\ref{sec:lumfun}). We use this
    LF to compute the weights to apply to each of the
    observed galaxies to take into account the unobserved ones
    (Section~\ref{sec:weights}).
\end{enumerate}
In future work we will update the estimated distances for the galaxies
using reconstructed velocity field and re-execute step (viii) to
update the corrections. The detail of this procedure will be discussed
in a later paper (Lavaux \& Hudson, 2011, in preparation).


\begin{figure*}
  \includegraphics[width=\hsize]{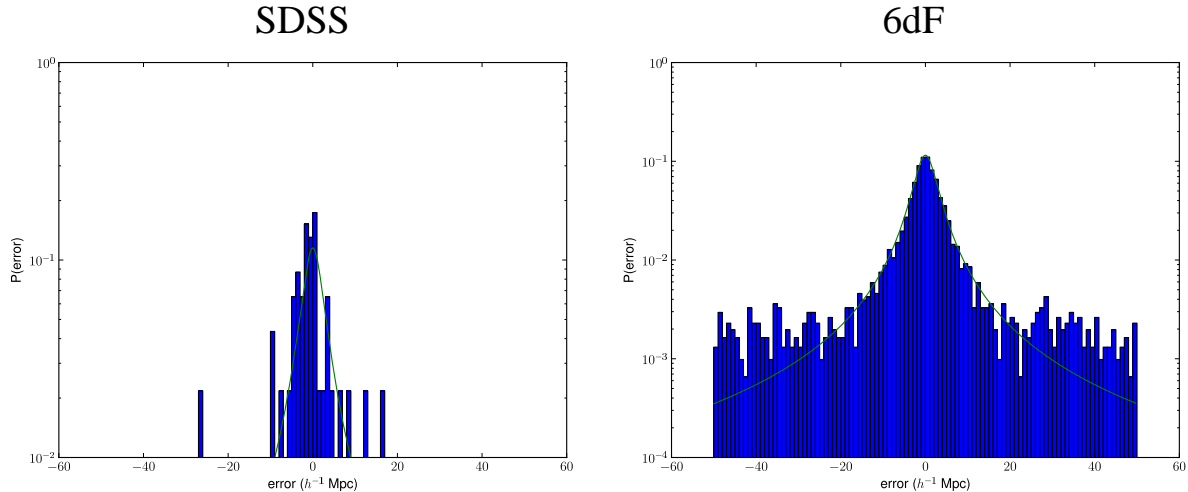}
  \caption{\label{fig:cloning_test} {\it Error distribution due to the
      redshift cloning procedure} -- We give here the computed error
    of the redshifts of either SDSS or 6dF redshift catalogue. We
    removed the redshift information of half of the objects in these
    catalogues and tried to recover them using the cloning
    procedure. The difference is plotted as an histogram in the two
    plots above. The overlaid continuous curve correspond to a Cauchy-Lorentz
    distribution with a width equal to $2.7$\Mpch. }
\end{figure*}

\subsection{Apparent magnitude corrections}
\label{sec:magcor}

We describe in this Section the corrections that must be applied to
apparent magnitudes to mitigate the effects of cosmological surface
brightness dimming, Galactic extinction and stellar evolution.  We
choose to use a definition of the magnitudes for target selection that
is related to the one used for defining magnitudes in the 2MRS. This
ensures that the final completeness is maximized in in the parts of
the sky where only redshifts from 2MRS are available.

The absolute magnitude $M$ at redshift zero of a galaxy may be written
as
\begin{equation}  
  M = m - A_K(l,b) - k(z) + e(z) - D_L(z)
\end{equation}
with $m$ the apparent magnitude, $A_K(l,b)$ the absorption by Milky
Way's dust in the direction $(l,b)$, $k(z)$ the $k$-correction due to
the redshifting of the spectrum, $e(z)$ is the correction for  evolution of the stellar population and $D_{L}$ is the luminosity distance. We convert the redshifts into luminosity distances assuming a $\Lambda$CDM cosmology with a mean total matter density parameter $\Omega_\text{M}=0.30$, and a Dark Energy density parameter $\Omega_\Lambda=0.70$. All absolute magnitudes are computed assuming $H_0=100$~\kmsMpc.

The absorption in $K_\text{S}$ band is related to the extinction $E_{B-V}$ estimated using the maps of \cite{Schlegel98} by the relation
\begin{equation}
  A_K(l,b) = 0.35 E_{(B-V)}(l,b)
\end{equation}
where the constant of proportionality is obtained from the relation
between absorption in $K$ band and in $V$ band \citep{Cardelli89}.

The adopted $k$-correction is
\begin{equation}
  k(z) = -2.1 z
\end{equation}
from \cite{BelMcIKat03}. Finally, the evolutionary correction is
\begin{equation}
  e(z) = 0.8 z
\end{equation}
also from \cite{BelMcIKat03}.

The magnitudes adopted in this work are circular aperture magnitudes defined
within a limiting surface brightness of 20 mag per square
arcsec. However, various effects will cause not only the observed
magnitude to change, but also the observed surface brightness. As the
surface brightness of the galaxy profile drops, the isophotal aperture
will move inwards and so the aperture magnitude will drop.

The surface brightness will depend on redshift via the usual $(1+z)^4$ cosmological
dimming effect as well as the extinction, $k-$correction and
evolutionary effects described above. Therefore the correction is
\begin{eqnarray}  
  \Delta \text{SB} & =  & \text{SB}(z=0) - \text{SB}(z) \nonumber \\
                         &  =  &  -10 \log(1+z) -A_K(l,b) - k(z) +  e(z) \,.
\end{eqnarray}
Note that the $k+e$ corrections have opposite sign to the cosmological
surface brightness dimming, and so there is some cancellation of these
effects. However, the cosmological term still dominates, so the net
effect is that as galaxies are moved to higher redshift their surface brightness is
dimmer.

By simulating simple S\'{e}rsic profiles, we have estimated how much the
aperture magnitude changes as a result of surface brightness 
dimming. For a typical 2M++ galaxy with  S\'{e}rsic  index $n = 1.5$ and
mean surface brightness within the effective radius of $\langle \mu_e \rangle = 17.5$,
we find that the correction to the magnitude due \emph{only} to a
change in aperture radius can be approximated by $0.16 \Delta
\text{SB}$ where $\Delta {SB}$ is the correction in the surface brightness. This term is
only the shift in magnitude due to the shift in isophotal radius, and
does not include the ``direct'' effect on the magnitude itself due to extinction
and $k+e$ corrections.  Thus the total effect is:
\begin{multline}
 \ktmpp = K_{20,c} + 1.16 [- A_K(l,b) - k(z) + e(z)] \\ -0.16 [10\log(1+z)]
  \label{eq:ktmpp_def}
\end{multline}
Note that this is close, but not identical, to the 2MRS corrected magnitude.

In some cases, only the magnitude $K_{20,e}$, derived from adjusting an ellipsoidal S\'{e}rsic profile, is available. In those cases, we have computed the corresponding $K_{20,c}$ using the following relation, obtained by fitting on the galaxies for which the two magnitudes were available:
\begin{equation}
	K_{20,c} = (0.9774 \pm 0.0005) K_{20,e} + (0.288 \pm 0.006). \label{eq:k20c_fit}
\end{equation}
The residual of the fit has a standard deviation equal to 0.11. We also use this relation whenever the predicted $K_{20,c}$ and the actual $K_{20,c}$ from 2MASS-XSC differs by 0.22 and calculate the $K_{20,c}$ from $K_{20,e}$. We have used this relation for 7\% of galaxies, both in the target and the final redshift compilation.

\subsection{Redshift cloning}
\label{sec:red_cloning}

Within the 6dF and SDSS regions, there is small-scale incompleteness due primarily to fibre collisions. 
To improve the redshift coverage of the catalogues we ``clone'' redshifts of nearby galaxies within each survey region. This procedure, which is related to another one described in \cite{NYUVAGC}, is as follows. Consider two targets $T_a$ and $T_b$. If $T_a$ does not have a measured redshift and $T_b$ has
one, and furthermore $T_b$ is the nearest target of $T_a$ with a
angular distance less than $\epsilon$, we copy the redshift of $T_b$
to $T_a$. $\epsilon$ is determined by the angular distance between two
fibres of the measuring instrument, which is $\epsilon=5'.7$ for 6dF
\citep{6dfDR1} and $\epsilon=55''$ for SDSS \citep{Blanton03}. 
We refer to redshifts cloned in this way as ``fibre-clones''.

To assess the errors on redshifts for the fibre-clones, 
we randomly split the set of galaxies which have a measured redshifts
in two sets $\mathcal{S}_\text{keep}$ and $\mathcal{S}_\text{test}$. 
We mark the galaxies belonging to $\mathcal{S}_\text{test}$ as having no redshift. We then apply the
fibre cloning procedure to these galaxies.

The result of this test is shown in Figure~\ref{fig:cloning_test} for
both SDSS galaxies and 6dF galaxies. We note that the Cauchy-Lorentz
distribution with width $W=2.7$\Mpch{} is a good fit to the central part of the two distributions. We used the formula
\begin{equation}
   P(e) = \frac{1}{\pi W} \frac{1}{1 + (e/W)^2}
\end{equation}
for the modelled probability distribution function in both
panels. We checked that a Gaussian distribution manages only to fit
the central part of the distribution and is less adequate than a
Cauchy-Lorentz distribution. The fibre-clones are given
a redshift error of $9\times 10^{-4}$, which corresponds to $\sim 2.7$\Mpch{} at redshift $z=0$.


\subsection{Redshift survey masks and completeness}
\label{sec:target_mask}

Because of the different redshift catalogues used in 2M++, we will separate the full sky into the following regions: $\ktmpp \le 11.5$ (2MRS); $\ktmpp \le 12.5$ (6dF/SDSS) or regions with insufficient redshift data. We begin by assigning preliminary ``masks'', then measure the redshift completeness and then assign final masks based on a completeness limit of 50\%. We now describe these steps in more detail.

\subsubsection{Preliminary mask selection}

The preliminary mask for the 6dF is as given in \cite{6dfDR1} : $|b| > 10^\circ$ and in the southern hemisphere $\delta \le 0^\circ$.

For SDSS, from the full DR7, we select only the most homogeneous and contiguous portion of the redshift survey. To obtain the geometry corresponding to this portion, we use the mask computed numerically from the MANGLE file, and impose an additional constraint on the positions: we keep only galaxies within the region $90^\circ < \alpha < 250^\circ$, to which we add another region at $250^\circ \le \alpha < 270^\circ$ and $\delta < 50^\circ$. This selection retains the major contiguous piece of the SDSS in the Northern Galactic cap, while removing the Southern Galactic strips and the small disjoint piece with $\alpha \sim 260$. After these cuts, we still retain 90\% of the area covered by the complete SDSS-DR7 ``Legacy'' spectroscopic survey.

\begin{figure}  
  \includegraphics[width=\hsize]{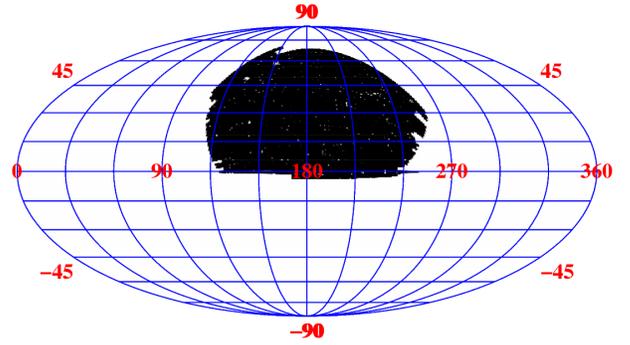}
  \caption{\label{fig:sdss_target_mask}
    Molleweide projection of the SDSS spectroscopic mask in Equatorial
    coordinates. We removed unconnected parts from the original
    mask. Note the presence of small holes in the mask due to both the
    presence of stars and not exact reconnection of the SDSS
    plates. This corresponds to an intersection of the geometry
    described by {\sc lss\_geometry.ply} and our selection criterion
    described in Section~\ref{sec:target_mask}. }
\end{figure}

The preliminary SDSS mask is shown in Fig.~\ref{fig:sdss_target_mask}. This mask has a relatively simple geometry and is contiguous, except for the presence of very small holes that are due either to stars or to imperfect overlap of the SDSS plates. These imperfections represent about 8\% of the
SDSS surface area.

For 2MRS, the initial mask is the whole sky with $|b|>5^\circ$, except
in the region $-30^\circ < l < +30^\circ$, where the Galactic latitude is $|b|>10^\circ$ \citep{EH06}, and excluding the regions covered by 6dF or SDSS. We refer to this region as 2Mx6S.

\begin{figure}
  \includegraphics[width=\hsize]{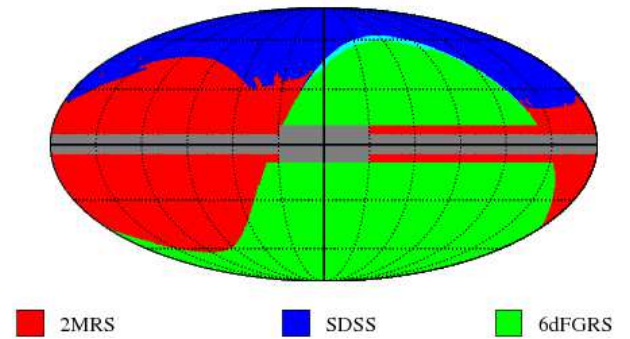}
  \caption{\label{fig:combined_coverage} A Molleweide projection in galactic coordinates showing the preliminary masks corresponding to the different redshift catalogues: the 2MASS Redshift Survey (red); the 6dF galaxy redshift survey (green), and the Sloan Digital Sky Survey DR7 (blue). Regions with no redshift data are shown in grey.}
\end{figure}

The combination of the three survey masks is shown in Fig.~\ref{fig:combined_coverage}, in Galactic coordinates. The grey area near the galactic plane is covered by none of the surveys. There is a small overlap between the SDSS and the 6dFGRS in the north galactic cap (in cyan).

\subsubsection{Redshift completeness}
\label{sec:red_completeness}

\begin{figure*}
  \includegraphics[width=\hsize]{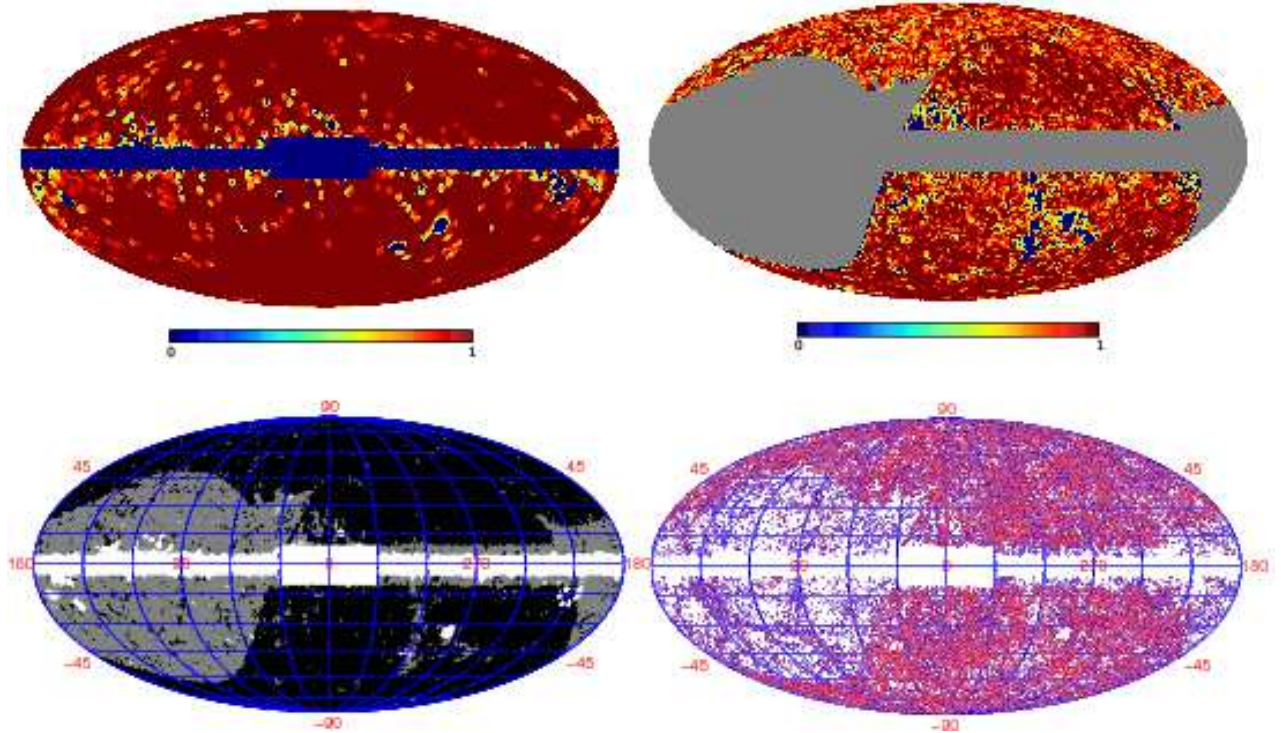}
  \caption{\label{fig:combined_masks} The 2M++ compilation in Galactic coordinates. The top panels give the redshift completeness of 2M++ for the limiting magnitude $\ktmpp=11.5$ (left panel) and $\ktmpp=12.5$ (right panel). The grey area corresponds to regions with no targets and no redshifts.
    In the bottom left panel, we show the final combined mask.
   The regions in white have been to completely excluded because of very low completeness. 
   The regions in grey (black respectively) have a redshift
    completeness higher than 50\% at a limiting magnitude  $\ktmpp\le
    11.5$ ($\ktmpp \le 12.5$ respectively). In the bottom right panel, we show all galaxies with redshift included in the
    2M++ compilation. Each galaxy is color-coded according to its redshift distance, blue for the nearest and red for the farthest.}
\end{figure*}

In order to determine LFs and weights needed for the density field, it is first necessary to assess the completeness of the redshift catalogues on the sky. We will estimate the redshift completeness in some direction of the sky for two different magnitudes cuts: $\ktmpp \le 11.5$ (for the 2Mx6S region) and $\ktmpp \le 12.5$ (for 6dF/SDSS regions).  The maps of these completeness are given in Fig.~\ref{fig:combined_masks}.  The completeness of the 2MRS is quite homogeneous and only drops close to the Galactic plane. The 6dF survey is mostly homogeneous except at the location of the Magellanic Clouds. The SDSS completeness is quite homogeneous, and remains at a level of about 80\% in the whole contiguous region.

\begin{table*}
	\caption{\label{tab:mask} Summary statistics for the primary regions in the 2M++ compilation}
	\begin{center}
		\begin{tabular}{lrrrrr}
		\hline
		Region & $m_{lim}$ & Area &  $N_{m<m_{lim}}$ & $N_{z}$ & $\bar{c}$ \\ \hline \hline 
		2Mx6S & 11.5 & 13 069 & 9 419  &  9 016 & 0.96 \\
		6dF  & 12.5  & 17 041 & 46 734 & 42 442 & 0.91 \\
		SDSS & 12.5  & 6 970  & 20 333 & 17 702 & 0.87 \\
		None &  --   & 4 172  & --    & --    & --   \\ \hline
	 {\bf Total} & --    & 37 080 & 76 451 & 69 160 & 0.90 \\ \hline
		\end{tabular}
	\end{center}
	Note that the regions are counted exclusively. We have not enforced the sample to have a local redshift 
	completeness higher than 50\%, resulting in a total number of redshifts higher than in the final catalogue.
\end{table*}

\subsubsection{Final masks and the 2M++ catalogue}

Based on the above analysis, we reject from 6dF and SDSS those regions where the completeness at $\ktmpp \le 12.5$ is less than 50\%, and assign these areas to the 2Mx6S and limit the magnitude there to 11.5. As a result, we note that while the 6dF and SDSS masks are contiguous, the 2Mx6S mask is not. We show in the lower left panel of Fig.~\ref{fig:combined_masks} the mask corresponding to the footprint of the 2M++ compilation. The black corresponds to regions where the survey has a completeness higher than 50\% at the limiting magnitude of $\ktmpp=12.5$ (the final 6dF and SDSS masks). The grey area represents the same but for a limiting magnitude of $\ktmpp=11.5$ (the final 2Mx6S mask). In white, we show the  parts of the sky where either there is no redshift information or target galaxies were not present.
 
The Zone of Avoidance is clearly visible. There are also important unobserved patches at $\ktmpp \le 12.5$ in the southern Galactic hemisphere at the locations of the Magellanic Clouds. The other white patches in the southern hemisphere are mostly related to local higher absorption by dust in the  Milky Way.

Thus, in summary, the final 2M++ catalogue is defined as all galaxies in 2Mx6S with $\ktmpp \le 11.5$, or in 6dF or SDSS with $\ktmpp \le 12.5$, and contains 
69~160
galaxies with redshifts (including fibre-clones). Table~\ref{tab:mask} summarizes the statistics and completeness for the different regions of the 2M++ compilation. We note that the 2M++ compilation redshifts are nearly 90\% complete, and so redshift completeness corrections are small.


\subsection{Redshift number density of galaxies}
\label{sec:red_number}

\begin{figure}
  \includegraphics[width=\hsize]{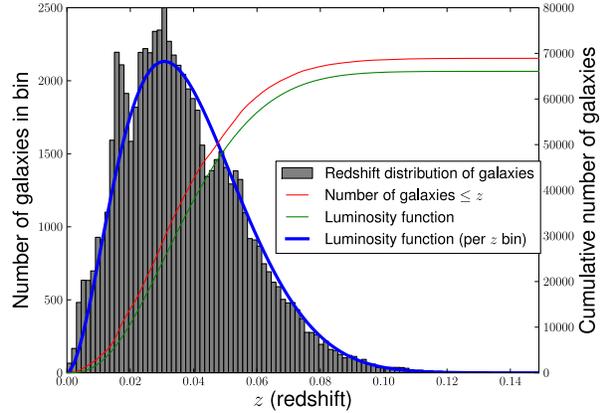}
  \caption{\label{fig:numcount} Redshift distribution of 2M++ galaxies. We show, in grey histogram, the number of galaxies within each redshift bin $\delta z = 0.00150$. The cumulative number of galaxies at a redshift less or equal to $z$ is given by the red curve. The predicted number of galaxies given by our fiducial LF given at the end of Section~\ref{sec:lumfun}, is shown in solid green for the cumulative number and in solid blue for the number of galaxies in each bin of the grey histogram. The LF has been fit using a subset of the catalogue for which $5,000$~\kms{}$ \le c z \le $20,000~\kms{} and $-25 \le M \le -21$. }
\end{figure}

Within the three regions outlined above, there are a total of 
69~160 galaxy redshifts (including fibre-clones). Fig.~\ref{fig:numcount} shows a histogram of all redshifts, as well as the cumulative counts starting from redshift $z=0$. Conservatively, the catalogue appears totally complete up to $z=0.02$ ($\sim 60$\Mpch{}). This is due to our use of the 2MRS for one part of the sky.

\begin{figure}
  \includegraphics[width=\hsize]{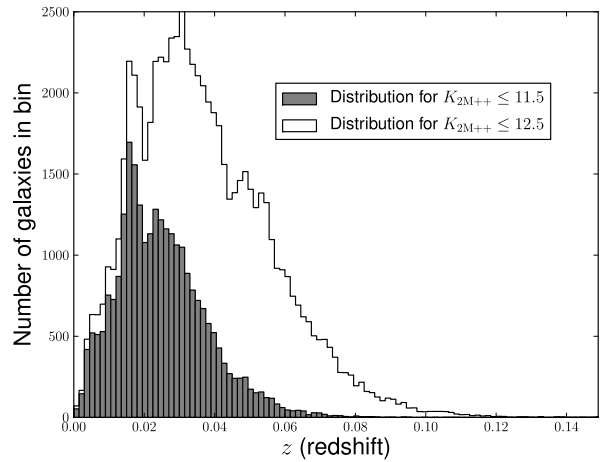}
  \caption{\label{fig:numcount_12_5}. Redshift galaxy distribution
      in 2M++ for $\ktmpp \le 11.5$ (filled histogram) and $\ktmpp \le 12.5$ (unfilled histogram).}
\end{figure}

In Fig.~\ref{fig:numcount_12_5}, we compare quantitatively the counts in the 2MRS region with $\ktmpp \le 11.5$ with the 6dF/SDSS regions with $\ktmpp \le 12.5$. Because our magnitude corrections are not precisely equivalent to those used for the 2MRS catalogue, the increase of the magnitude cut to $\ktmpp=12.5$ is not strictly equivalent to using only the 6dF and SDSS spectroscopic data but also includes a few 2MRS galaxies. Nonetheless, the increase in the magnitude cut correspond mostly to the the sky portions covered by both the SDSS and the 6dF. The deeper redshift data allows us to better probe more distant large-scale structures, particularly in the redshift $0.02 \la z\la 0.05$. For example, the feature in the redshift distribution at $z \sim 0.05$ corresponding to the Shapley Concentration is not present for the subcatalogue $\ktmpp \le 11.5$, while it is clearly seen in the 6dF subcatalogue ($\ktmpp \le 12.5$).

\subsection{Luminosity function}
\label{sec:lumfun}

\begin{figure*}
  \includegraphics[width=\hsize]{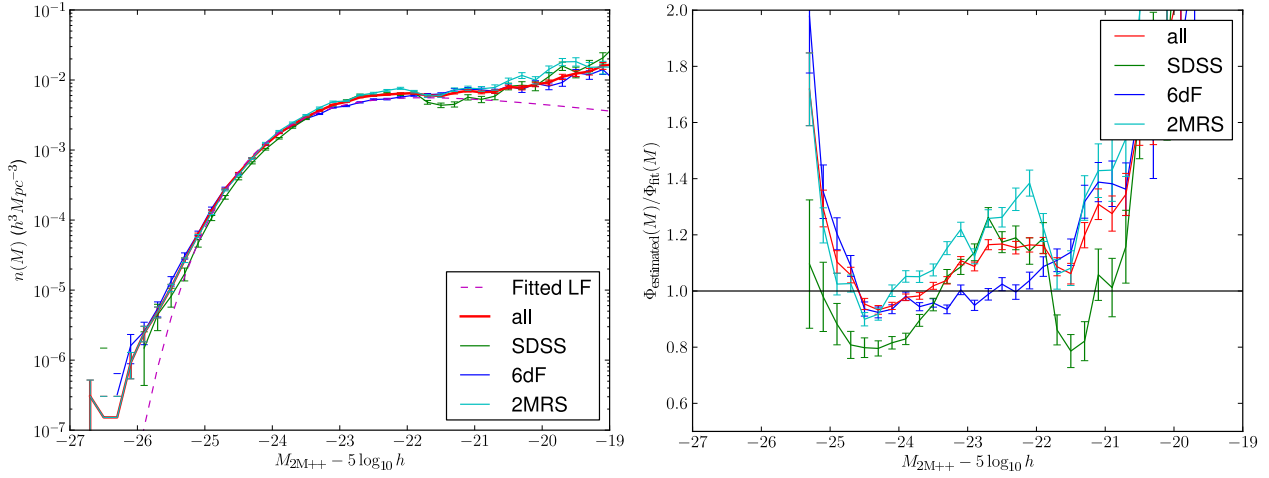}
  \caption{\label{fig:lf_fun} Galaxy LF estimates. The left-hand panel shows the non-parametric LF estimated using the $1/V_\text{max}$ method is shown by the solid lines for the regions covered either by 2MRS, 6dF, SDSS or all together. For these plots we use data from 750~\kms{} to 20~000\kms{}. The dashed line shows the parametric LF using the likelihood method of Section~\ref{sec:lumfun} for galaxies with absolute magnitudes in the range $-25 \le M \le -21$, for redshift distances 5~000~\kms{} to 20~000~\kms{}. The error bars reflect only the uncertainties in galaxy counts and do not include cosmic variance effects. The right-hand panel shows the difference between the $1/V_\text{max}$ LFs and the fitted parametric LF.} 
\end{figure*}

\subsubsection{Method}

In order to correct for selection effects due to magnitude limits, it is first necessary to measure the LF.
We take into account the redshift completeness to measure the LF of galaxies in the combined catalogue. We use a modified version of the likelihood formalism used to find \cite{Sch76} function parameters, as described by \cite{Sandage79}.  We assume that evolutionary effects on the luminosity of galaxies have been accounted for by Eq.~\eqref{eq:ktmpp_def}.
The parametrization adopted is the usual Schechter function:
\begin{equation}
  \Phi(L) = \frac{n^*}{L^*} \left(\frac{L}{L^*}\right)^{\alpha}
  \exp\left(-\frac{L}{L^*}\right),
\end{equation}
with $n^*$ the density normalization constant, $L^*$ the
characteristic luminosity break, or equivalently in terms of absolute
magnitudes
\begin{multline}
  \Phi(M) = \\ 0.4 \log(10) n^* 10^{0.4(1+\alpha)(M^*-M)}
  \exp\left(-10^{0.4(M^*-M)}\right) \\= n^* \Phi_0(M),
\end{multline}
with $M^*$ the characteristic absolute magnitude break in the
Schechter function. Above, we have introduced $\Phi_0(M)$, which is the unnormalized
Schechter function. We model the probability of observing a galaxy of
absolute magnitude $M_i$ given its redshift $z_i$ as
\begin{multline}
   P(M_i|z_i,\alpha,M^*,n^*,c) = \\ \frac{c(M_i,\hat{u}_i,d_i)
     \Phi_0(M_i)}{\int^{M_\text{max}}_{M_\text{min}} c(M,\hat{u}_i,d_i)
     \Phi_0(M)\,\text{d}M} \label{eq:P_one_galaxy}
\end{multline}
with $M_\text{min}$, $M_\text{max}$ the maximum absolute magnitude range from which the galaxies were selected in the catalogue, $c(M,\hat{u}_i,d_i)$ the completeness in the direction
$\hat{u}_i$ of the object $i$, at the absolute magnitude $M$, $d_i$
the luminosity distance of the galaxy $i$ at redshift $z_i$.  This
expression is simplified using our assumption that redshift
incompleteness $c(M,\hat{u},r)$ may be modelled by two maps at two
apparent magnitude cuts. $c(M,\hat{u},r)$ is thus
\begin{equation}
  c(M,\hat{u},r) = \left\{
      \begin{array}{ll}
        c_b(\hat{u}) & \text{if } M + 5
        \log_{10}\left(\frac{r}{10\text{ pc}}\right) \le m_b
        \\ c_f(\hat{u}) & \text{if } m_b < M + 5
        \log_{10}\left(\frac{r}{10\text{ pc}}\right) \le m_f \\ 0 &
        \text{otherwise,}
      \end{array}
      \right.
\end{equation}
with $m_b=11.5$ and $m_f=12.5$.  The expression of the probability
\eqref{eq:P_one_galaxy} may thus be newly expressed as 
\begin{equation}
  P(M_i|z_i,\alpha,M^*,n^*,c) = \frac{c(M_i,\hat{u}_i,d_i)
    \Phi_0(M_i)}{f(d_i,\hat{u}_i,M_\text{min},M_\text{max})}
\end{equation}
with
\begin{multline}
  f(r,\hat{r},M_\text{min},M_\text{max}) = c_b(\hat{r})
  \Gamma^{M^*,\alpha}_{M_\text{min},M_\text{max}}(m_b,r_{10}) + \\ c_f(\hat{r})
  \left(\Gamma^{M^*,\alpha}_{M_\text{min},M_\text{max}}(m_f,r_{10})-\Gamma^{M^*,\alpha}_{M_\text{min},M_\text{max}}(m_b,r_{10})\right),
\end{multline}
the normalization coefficient for the direction $\hat{r}$ at distance $r$, and $r_{10}$ defined as the distance in units of 10 pc.
In the above, we have also used the function $\Gamma^{M^*,\alpha}_{M_\text{min},M_\text{max}}$ defined as
\begin{multline}
  \Gamma^{M^*,\alpha}_{M_\text{min},M_\text{max}}(m,r_\text{10}) = \\
  	\Gamma_\text{inc}\left(1+\alpha,10^{0.4 \left(  M^* - \min(\max(M(m,r_{10}),M_\text{min}),M_\text{max})\right)}\right)
\end{multline}
with the absolute magnitude
\begin{equation}
	M(m,r_{10}) = m - 5\log_{10}(r_{10})
\end{equation}
and $\Gamma_\text{inc}(a,y)$ the incomplete Gamma function
\begin{equation}
   \Gamma_\text{inc}(a,y) = \int_y^\infty x^{a-1} \text{e}^{-x}\;\text{d}x\,.
\end{equation}

We write the total probability of observing the galaxies with
intrinsic magnitude $\{M_i\}$ and redshift $\{ z_i \}$ given the
Schechter LF parameters:
\begin{multline}
   P(\{M_i\}| \{c_i\}, \{z_i\},\alpha,M^*,n^*) = \\ 
     \prod_{i=1}^{N_\text{galaxies}} P(M_i|z_i,\alpha,M^*,n^*,c_i).
\end{multline}
Using Bayes theorem, we now estimate the most likely value taken by
$\alpha$,$M^*$ assuming a flat prior on these parameters.

The normalization constant $n^*$ is determined using the minimum variance estimator of \cite{DavisHuchra82}, but neglecting the effects of cosmic variance on the weights by setting $J_3=0$. While our estimate may be biased relative to the galaxy mean density outside the catalog, it is less noisy than the optimal case. The estimate also corresponds better to the density in the piece of Universe that we consider than the density corresponding to the optimal weighing. Our choice leads also to a simplification of the mean density as the total number of galaxies divided by the effective volume of 2M++. Consequently, for a survey limited in the absolute magnitude range $[M_\text{min},M_\text{max}]$ and with volume $V$ we compute the mean density of galaxy $\bar{n}$ by the following equation
\begin{equation}
	\bar{n} = \frac{N_\text{galaxies}}{\int_V\text{d}^3{\bf r} f(r,\hat{r},M_\text{min},M_\text{max})}, \label{eq:nbar}
\end{equation}
with a standard deviation only from Poisson noise
\begin{equation}
	\frac{\sigma_{\bar{n}}}{\bar{n}} = \frac{\sqrt{N_\text{galaxies}}}{N_\text{galaxies}}. \label{eq:err_nbar}
\end{equation}
because we have set $J_3 = 0$.
We then convert $\bar{n}$ into $n^*$ using
\begin{equation}
	n^* = \frac{\bar{n}}{\int_{M_\text{min}}^{M_\text{max}} \Phi_0(M)\,\text{d}M}.
\end{equation}
Similarly it is possible to define the luminosity density
\begin{equation}
  \bar{L} = n^* 10^{0.4(\mathrm{M}_\odot - M^*)} \Gamma(2+\alpha) \times (1 \mathrm{L}_\odot), \label{eq:lbar}
\end{equation}
with $\Gamma(a) = \Gamma_\text{inc}(a,0)$. The luminosity density is less sensitive than the number density to fluctuations in $\alpha$.

To determine the LF parameters, we select a subset of the galaxies in our catalogue. We have defined the subset by the joint conditions:
\begin{itemize}
  \item[-] Galaxies must have a redshift $z$ such that 5,000~\kms{}$\le c z \le\,$20,000~\kms{}. The lower limit reduces the impact of peculiar velocities on absolute magnitude estimation, which is derived using redshifts in the CMB rest frame. By limiting to $c z \le 20,000$~\kms{}, we avoid more distant volumes with high incompleteness.
  \item[-] The absolute magnitude estimated from the redshift in CMB rest frame is within the range $[M_\text{min}=-25,M_\text{max}=-21]$. As mentioned later in this Section, this magnitude selection removes the bright objects that do not seem to follow a Schechter LF \citep[as also discussed][]{JPCS06}.
\end{itemize}
Absolute magnitudes are determined with $H = 100h$~\kmsMpc with $h=1$ and we have assumed a flat $\Lambda$CDMcosmology $\Omega_\text{m}=0.30$ and $\Omega_\Lambda=0.70$. 
We do not distinguish between early-type and late-type galaxies, and so fit both populations with a single parameter.  

\subsubsection{Results}

The derived LF parameters are summarized in Table~\ref{tab:lf} for our default choice of cuts discussed above as well as for other choices that we discuss below. The error-bars are given at 68\% confidence limit, estimated using a Monte-Carlo-Markov-Chain method. 

Fig.~\ref{fig:lf_fun} shows the LF for our default cuts in the CMB rest frame. We also show, for the entire data set and for each subcatalogue, the non-parametric LFs estimated using the unbiased $1/V_\text{max}$ method \citep{S68,F76}. Note that for the $1/V_\text{max}$ LFs the volume and magnitude limits are different than for the parametric fit, which explains that the fitted parametric faint-end slope is not a good fit to the $1/V_\text{max}$. in the range $[-21,-19]$. In the left panel, we give the LFs and in the right panel the ratio between the estimated LFs and the best fit.  
\begin{table*}
  \begin{center}
  \caption{\label{tab:lf} Summary of K-band Schechter LF parameters from this paper and selected results from the literature. Magnitude ranges
  with a $\sim$ are estimated.  $n_*$ is in units of $(10^{-2}
  h^3\text{ Mpc}^{-3})$ and the luminosity density $\bar{L}$ is in units
  of $10^8 h\; L_{\sun} \text{Mpc}^{-3}$, assuming $\mathrm{M}_{K,\odot} = 3.29$}
  \begin{tabular}{llllllll}
    \hline
    Reference & Frame & Magnitude & Redshift range & $\alpha$ & $M^* -
    5 \log_{10} h$ & $n_*$ & $\bar{L}$\\
    \hline
    \cite{Kochanek01} & & $[-26;-20]$ &  $[2 000;14 000]$  & $-1.09 \pm 0.06$ & $-23.39\pm 0.05$  & $1.16\pm 0.10$ \\
    \cite{Cole01}     & & $\sim[-26;-20]$  &       ?           & $-0.96\pm 0.05$ & $-23.44\pm 0.03$      & $1.08\pm 0.16$ \\
    \cite{Bell03}     & & $\sim[-25;-18]$  &         ?           & $-0.77\pm 0.04$ & $-23.29\pm 0.05$      & $1.43\pm 0.07$ \\
    \cite{Eke05}      & & $\sim[-25;-20]$  &        ?          & $-0.81\pm 0.07$ & $-23.43\pm 0.04$      & $1.43\pm 0.08$ \\
    \cite{2MRS}       & & $\sim[-28.5;-16]$     &   ?         & $-1.02$         & $-23.4$              & $1.08$ \\
    \cite{JPCS06}     & & $[-28.85;-15.5]$ &  $[750;+\infty]$  & $-1.16\pm 0.04$ & $-23.83\pm 0.03$      & $0.75\pm 0.08$ \\
    \hline
    {\bf This work} & CMB & $[-25;-21]$ & $[5 000; 20 000]$  & $-0.73
    \pm 0.02$ & $-23.17 \pm 0.02$ & $1.11 \pm 0.02$ & $3.94\pm 0.02$\\
    This work & CMB &  $[-25;-17]$ & $[750; 20 000]$  & $-0.80 \pm 0.01$ & $-23.22 \pm 0.01$ & $1.13 \pm 0.02$ & $4.16\pm 0.02$\\
    This work & LG & $[-25;-17]$ & $[750; 20 000]$ & $-0.86 \pm 0.01$& $-23.24 \pm 0.01$ & $1.13 \pm 0.02$ & $4.25 \pm 0.02$\\
    This work & LG  & $[-25;-21]$ & $[5 000; 20 000]$ & $-0.76 \pm 0.02$ & $-23.18 \pm 0.01$ & $1.14 \pm 0.02$ & $4.02 \pm 0.02$ \\
    $|b|>10$, $K<11.5$ & CMB &  $[-25;-17]$ & $[300; 20 000]$  & $-0.94 \pm 0.02$ & $-23.28 \pm 0.01$ &  \\
    $1/V_\mathrm{max}$ fit & CMB & $[-25;-21]$ & $[750; 20 000]$ & $-1.03 \pm 0.02$ & $-23.43 \pm 0.01$ & $0.85 \pm 0.06$ & $4.22\pm0.11$ \\
    \hline 
  \end{tabular}
  \end{center}
\end{table*}

\subsubsection{Discussion and comparison with previous results}

Table~\ref{tab:lf} also lists LF parameters from previous 2MASS studies. Our fitted LF parameters are in agreement with previous studies of the K-band LF \citep{Bell03,Eke05} but are somewhat different than those found by \cite{Kochanek01}, \cite{Cole01}, \cite{2MRS} and \cite{JPCS06}.

The derived LF parameters are sensitive to a number of systematic effects: the magnitude range used, the rest-frame used for the redshifts, and the fitting method itself.

The Schechter function itself appears not to perfect fit over the whole range of magnitudes. Consequently, the fitted parameters depend in the magnitude (and distance) range of the galaxies used in the fit. Our default minimum distance $r \geq 5000$~\kms{} corresponds to $M_K \lesssim -21$ for $\ktmpp = 12.5$.  However, the $1/V_\text{max}$ method seems to indicate an inflection in the LF at $M_K \sim -21$. This bend is also seen by \cite{Bell03} and \cite{Eke05}. Indeed \cite{Bell03} attempted to fit the part at $M_K > -21$ with a power-law instead of a Schechter function.) Several studies \citep{BivDurGer95,YagKasSek02} have noted a dip in the LF of cluster galaxies at a similar location (approximately 2 magnitudes below $M^*$), although other studies suggest that it is a flattening rather than a dip \citep{Tre98}.  In any case, it seems clear that the choice of magnitude range will affect the Schechter LF parameters.  In \cite{JPCS06} and \cite{Cole01}, the magnitude range used in the fit is fainter than our default. 

A second issue, which arises when using galaxies with very low redshifts, is the choice of flow model or rest-frame redshifts. Very nearby galaxies are likely to share the peculiar velocity of the Local Group (LG), so the redshift in the LG frame is a better proxy distance than the CMB-frame redshift. For better understanding of the dependence of our results on both local flows and clustering, we have fit the parameters of the Schechter function in two rest frames (CMB or LG). We find that, for samples extending to $M_K \sim -17$, the faint-end slope $\alpha$ steepens, but only by 0.06.

Finally, the magnitudes, correct and the fitting method itself are probably the most important systematics. 
\begin{enumerate}

\item We note that the studies of \cite{Cole01}, \cite{Eke05} and
  \cite{Bell03} are based on Kron magnitudes, and that of
  \cite{JPCS06} is based on total magnitudes, leading to a possible
  difference in $M^*$ of $0.20\pm 0.04$, as discussed by
  \cite{Kochanek01}. 

\item Another notable difference is that \cite{JPCS06} have tried to
  integrate the effect of uncertainties on the determination of
  magnitudes, which we do not do here.

\item \cite{Bell03} matches SDSS redshifts to both the 2MASS XSC and
  the PSC catalogues.  \cite{Bell03} argues that selection effects
  bias the raw 2MASS LF compared to the true LF.  However, whereas
  those authors were interested in, for example, the total stellar
  mass density in the nearby Universe, \emph{our goal} is rather a
  consistent magnitude system coupled with uniform selection across
  the sky. Since our primary method will be to weight by luminosity,
  the small missing contribution from low surface brightness galaxies
  and the low surface brightness regions of catalogued galaxies is of
  little concern to us.

\item The fitting method itself may also make a difference.  Our
  default parametric fit is pinned to the magnitude range where the
  formal Poisson errors are smallest, namely $[-25,-22]$. However, we
  have seen that systematic effects can be important. As an
  alternative, we have taken the LF given by the $1/V_\mathrm{max}$
  method, added in quadrature the statistical error bars and the
  fluctuations from the different subcatalogues, and fitted these data
  with a Schechter LF. As indicated in Table~\ref{tab:lf}, we have
  obtained a steeper faint end slope and a brighter $M_*$, which are
  in better agreement with \cite{Kochanek01}, \cite{Cole01} and \cite{2MRS},
  but still discrepant with \cite{JPCS06}. 
\end{enumerate}

We conclude that, given all of these systematics, our LFs are reasonably consistent with those that have been found previously. One aspect which can be improved is peculiar velocity corrections, but we postpone a fully self-consistent treatment of peculiar velocities and the LF determination to a future paper.

We confirm that the bright end part of the LF does not seem to follow a Schechter LF,  as already seen by the 6dfGRS \citep{JPCS06}. This effect is clearly seen in the SDSS, 2MRS and 6dfGRS subsamples separately. The deviation becomes significant at $M_{K} \lesssim -25$, or two magnitudes brighter than $M_{*}$, and is presumably due to brightest cluster galaxies, which have typical $K$-band magnitudes of $\sim -26$ \citep{LinMoh04} and have long been known to deviate from the extrapolation of a Schechter function \cite{TreRic77}.

We may check the consistency of this LF with the number of galaxies in the 2M++ catalogue. We predict that the total number of galaxies of redshifts between the distances $r_\text{min}(z_\text{min})$ and $r_\text{max}(z_\text{max})$, assuming the Schechter LF, is
\begin{multline}
  N = \int_{M_\text{min}}^{M_\text{max}} \Phi(M)\text{d}M \int_{r_\text{min}}^{r_\text{max}} \text{d}^3{\bf r}\,f(r,\hat{r},M_\text{min},M_\text{max}) \label{eq:total_number}.
\end{multline}
We plot this function as a solid green line in Fig.~\ref{fig:numcount}. We also show the predicted number of galaxies in each bin of the grey histogram by a solid blue line. We see that the prediction in each redshift bin agrees well with the observed number of galaxies, but the total is off by $\sim$2\%. The difference comes both from the low luminosity part of the luminosity which is not adjusted because of our cut at $c z \ge 5,000$~\kms{} and the high luminosity part for which objects are not following a Schechter LF, as in Fig.~\ref{fig:lf_fun}.

In Table~\ref{tab:lf}, we also give the mean luminosity density $\bar{L}$ as derived from Eq.~\eqref{eq:lbar}. $\bar{L}$ is a lot less sensitive than $\bar{n}$ to the faint end of the luminosity function. 
As before, the errors are dominated by systematics due to the different corrections from peculiar velocities and the adequacy of the Schechter function to fit the observed luminosity function. Taking the average and computing the dispersion in values for $\bar{L}$ for the four tests indicated in Table~\ref{tab:lf} yields $\bar{L}=(4.09\pm 0.12) 10^8 h\; \mathrm{L}_\odot$.


\subsection{Weights}
\label{sec:weights}

Using the LF, we may now compute the appropriate weights to give to observed galaxies to account for incompleteness of the redshift catalogue. Our long-term goal is to reconstruct the dark matter density, under the assumption that galaxies trace the dark matter. There are several ways to link the galaxy density to the dark matter density: assuming that there is a linear relation between the two fields, one might consider number-weighting, in which the DM density is assumed to be related to the number-density of galaxies, or luminosity-weighting, which can serve as proxy for stellar mass, and so may be a better tracer of DM density. We will consider both of these schemes here. More complicated relationships, for example based on a halo model \citep{MarHud02}, will be considered in a future paper.

We compute number-weighting based on the fraction of observed galaxies:
\begin{multline}
  f^N_\text{observed}({\bf r},M_\text{min},M_\text{max}) = \frac{N_\text{observed}({\bf r})}{N_\text{average}} \\ = 
       \frac{f(r,\hat{r},M_\text{min},M_\text{max})}{\int_{M_\text{min}}^{M_\text{max}} \Phi_0(M)\,\text{d}M}. \label{eq:numweight}
\end{multline}
The weight applied to each galaxy is then $1/f^N_\text{observed}({\bf r})$. This
procedure is common and has been used previously \citep[e.g.][]{DavisHuchra82,PikeHudson05}

We follow a similar procedure for correcting the local luminosity
density of galaxies by estimating how much light we are missing at the distance of each
galaxy located at position ${\bf r}$. The fraction
\begin{multline}
  f^L_\text{observed}({\bf r}) = \frac{L_\text{observed}({\bf
      r})}{L_\text{average}} = \\ \frac{1}{L_\text{average}}\left(
  (c_b-c_f)(\hat{r}) \Gamma^{M^*,1+\alpha}_{M_\text{min},M_\text{max}}\left(m_b,r_{10}\right) + \right. \\ \left. c_f(\hat{r})
  \Gamma^{M^*,1+\alpha}_{M_\text{min},M_\text{max}}\left(m_f,r_{10}\right)
  \right) \label{eq:lumweight}
\end{multline}
of luminosity with $L_\text{observed}$ is the mean luminosity expected
to be observed in a small volume at position ${\bf r}$, $\hat{r}={\bf
  r}/|{\bf r}|$ and $L_\text{average}$ the mean luminosity emitted by
galaxies in the Universe. The value of $L_\text{average}$ is
\begin{equation}
	L_\text{average} = \int_{M_\text{min}}^{M_\text{max}} L(M) \Phi_0(M)\,\text{d}M.
\end{equation}
The weight to apply to each intrinsic luminosity of a galaxy is then $1/f^L_\text{observed}({\bf r})$. This
procedure has already also been used with success with observation and mock catalogues
\citep[e.g.][]{Lavaux08,Lavaux10, DavNusMas10}.

For our choice of absolute magnitudes, $M_\text{min}=-25$ and $M_\text{max}=-20$, the 2M++ is volume limited up to $r_\text{min}\sim 20$\Mpch{}, and extends up to $r_\text{max}=$300\Mpch{}. We find that at a distance of $\sim 150$\Mpch{}, the galaxy number weights are typically between 10 and 400, depending on whether the region is limited to $\ktmpp\le 12.5$ or $\ktmpp \le 11.5$, respectively. Similarly, the luminosity weights range between $\sim$2 and $\sim$40. So weighing by luminosity has the advantage that it is less noisy at large distances.


\section{Treating the Zone of Avoidance}
\label{sec:zoa}

\begin{figure}
  \includegraphics[width=\hsize]{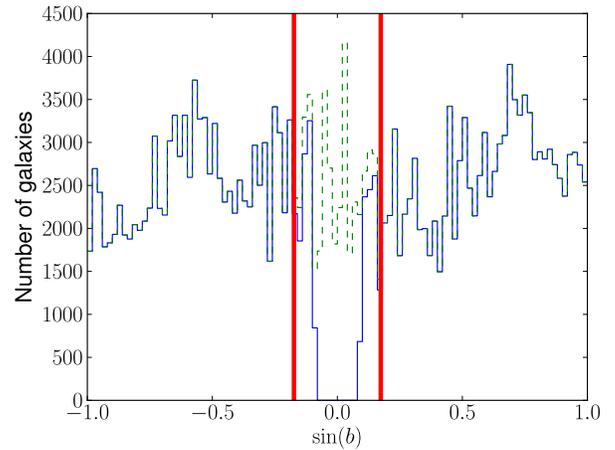}
  \caption{\label{fig:zoa} The effect of the ZoA on 2M++. The weighed number density of galaxies in each bin of $\sin(b)$ is shown by the  thin solid histogram. The dashed line shows  the number density of galaxies once ZoA is filled with cloned galaxies. The two thick vertical lines correspond to $b=\pm 10^\circ$. 
  Here we used only galaxies for which $c z \le 15,000$~\kms{}.}
\end{figure}

The ``Zone of Avoidance'' (ZoA) is the region of the Galactic plane where observations of galaxies are difficult due to the extinction by Galactic dust and stellar confusion. 
We show in Fig.~\ref{fig:zoa}, the number of galaxies in 2M++ with
$\ktmpp\le 11.5$, in bins of $\sin(b)$, and corrected for
incompleteness effects. We see that the distribution is close to flat as a function of Galactic latitude,
except for a hole contained between the latitudes $-10^\circ \le b \le
10^\circ$.  We define the ZoA in 2M++ as this band for galactic longitudes $-30 \leq l \leq 30$, but reduce it to $5^\circ$ outside this range. In addition, we impose the constraint that the absorption not to exceed $A_K=0.25$ in regions devoid of galaxies.

In order to reconstruct the density field over the full sky, it is clearly necessary to fill the ZoA.
One option is to fill it with mock galaxies so that their density of these objects matches the mean density outside the ZoA. 
This option, would however, fail to interpolate large-scale structure observed above and below the ZoA. 
The option adopted here, following \cite{LB89}, is to ``clone''  galaxies immediately above and below the ZoA. 
The procedure of creating a galaxy clone at a latitude $b_c$ of a galaxy at latitude
$b$ is simply to shift the latitude 
\begin{equation}
  \sin(b_c) = \sin(b_\text{zoa}) - \sin(b)
\end{equation}
where 
\begin{equation}
	b_\text{zoa} = \left\{\begin{array}{ll}
						\text{sign}(b) \times 5^\circ & \mathrm{if }\,|b| > 5^\circ\, \mathrm{ and }\, |l| > 30^\circ \\
						\text{sign}(b) \times 10^\circ & \mathrm{if }\,|b| > 10^\circ\, \mathrm{ and }\, |l| < 30^\circ \\
					\end{array}	\right.
\end{equation}
We refer to these as ``ZoA-clones''.
Fig.~\ref{fig:zoa} shows the distribution of galaxies as a function of Galactic latitude before and after cloning.
After cloning, the distribution shows no dependence on latitude.


\section{Grouping galaxies}
\label{sec:groups}

We use redshifts to estimate galaxy distances, but in the presence of peculiar velocites this relationship is not perfect. In addition to the so-called ``\citet{Kaiser87} effect'' which affects very large scales, there is also a contamination by the ``Finger-of-god effect'' due to the velocity dispersion of galaxies in clusters of galaxies.  This causes a significant amount of noise on the redshift-estimated distance. One way to deal with the problem is to group galaxies, which by simple averaging improves the distances estimated from the redshifts. The grouping information is also interesting to study the statistics and properties of galaxy groups.

In this Section, we describe the algorithm used to assign galaxies to
groups and clusters.  We use this information in the next sections for deriving a better density field (Section~\ref{sec:cosmography}) and, in a future work, peculiar velocities. Grouping also allows a better determination of the center of mass of superclusters (Section~\ref{sec:masses}) and their infall pattern (Section~\ref{sec:infall}). As a byproduct of 2M++, we provide the a catalog of groups and their properties in Appendix~\ref{app:2mpp_group}.


\subsection{Grouping Algorithm}

\begin{figure}
  \includegraphics[width=\hsize]{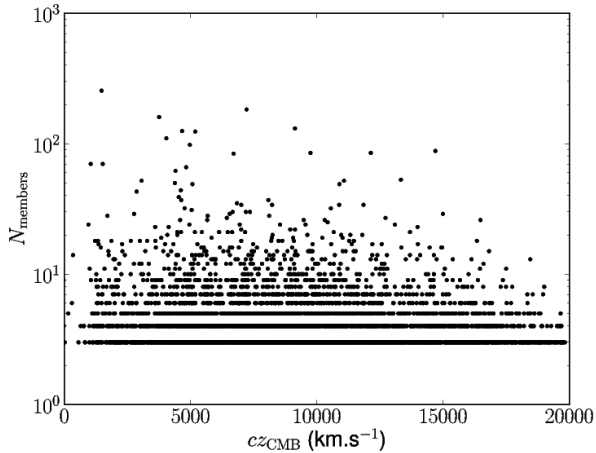}
  \caption{\label{fig:number_of_galaxies_in_groups} 
    Number of group members (richness) as a function of redshift. The richness is is not corrected for incompleteness.
    }
\end{figure}

\begin{figure}
  \includegraphics[width=\hsize]{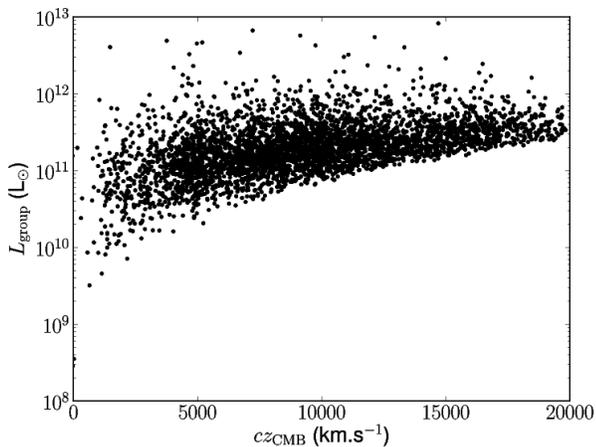}
  \caption{\label{fig:lum_in_groups} Group luminosity
      as a function of redshift. The luminosities are not corrected
    for incompleteness.}
\end{figure}

To assign galaxies to groups we use the standard percolation, or ``Friends-of-friends'' algorithm developed by \cite{HuchraGeller82}. The algorithm is designed to identify cone-like structures in redshift space. Two galaxies are considered to be part of the same group if:
\begin{itemize}
  \item[-] their estimated angular distance separation is less than
    $D_\text{sep}$,
  \item[-] their apparent total velocity separation separation is less
    than $V_0$.
\end{itemize}
$V_0$ is kept fixed for the whole volume of the
catalogue. $D_\text{sep}$ is adapted such that the detected structures
are always significant compared to the apparent local number density
of galaxies, by explicitly accounting for selection effects. 
The constraint of a constant local overdensity at a
redshift distance $z$ leads to
\begin{equation}
  D_\text{sep} = D_0 \left( \frac{\int_{L_\text{min}(z)}^{+\infty} \Phi(L)\,\text{d}L}{\int_{L_\text{min}(z_\text{F})}^{+\infty} \Phi(L)\,\text{d}L}\right)^{-1/3},
\end{equation}
with $L_\text{min}(z)$ the minimum absolute luminosity observable at
redshift $z$, $\Phi(L)$ the galaxy LF, $z_\text{F}$
the fiducial redshift, $D_0$ the selection angular distance at
fiducial redshift. The parameter $D_0$ is linked to the sought overdensity
$\delta_\text{overdensity}$ for group detection by the relation
\begin{equation}
  \delta_\text{overdensity} = \left(\frac{4\pi}{3} D_0^3
  \int_{L_\text{min}(z_\text{F})}^{+\infty}
  \Phi(L)\,\text{d}L\right)^{-1} - 1.
\end{equation}
This density is computed at the fiducial redshift distance $z_\text{F}$.

We have chosen the following parameters for defining our groups:
$V_\text{F} = c z_\text{F} = 1,000$~\kms{} and $\delta_\text{overdensity}=80$. These parameters have been used in previous studies \citep{Ramella89}.  With the LF, for our choice of fiducial parameters, we compute that the transverse linking length is $D_0=0.45$\Mpch{}. We count 4~002 groups with three or more members within 2M++, for redshift distance less than 20,000~\kms{}. We do not group galaxies farther than 20,000~\kms{} where the catalogue becomes sparse. For the very nearby Virgo and Fornax clusters, the FoF algorithm fails and so we manually assign galaxies to nearby clusters according to the parameters given in table~\ref{tab:manual_groups}. 

\begin{table}
  \begin{center}
    \begin{tabular}{cccccc}
      \hline
      Group name & $c z_\text{min}$ & $c z_\text{max}$ & $\theta_\text{sep}$ & $l$ & $b$ \\
      \hline
      \hline
      Virgo & $-\infty$ & 2,500~\kms{} & 10$^\circ$ & 279 & 74 \\
      Fornax & $-\infty$ & 1,600~\kms{} & 8$^\circ$ & 240 & -50 \\
      \hline
    \end{tabular}
    \caption{\label{tab:manual_groups} Parameters for manual grouping
      of galaxies. All galaxies which are in the direction $(l,b)$ and
      within the redshifts $[z_\text{min};z_\text{max}]$ with a
      maximum angular separation to $(l,b)$ equal to
      $\theta_\text{sep}$, are considered part of the group indicated
      in the first column.}
  \end{center}
\end{table}


\subsection{Results}

\begin{figure}
  \includegraphics[width=\hsize]{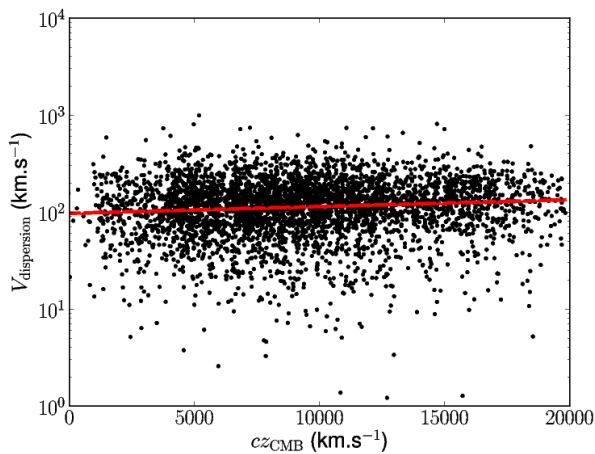}
  \caption{\label{fig:veldisp_in_groups} Group velocity dispersions
      as a function of redshift. The thick solid  line indicates the trend of the evolution of the average velocity dispersion with redshift.
    The scale of the variation is $\sim 1.4h^{-1}$~Gpc, far larger than the depth of the 2M++ catalogue.}
\end{figure}

In Figure~\ref{fig:number_of_galaxies_in_groups}, we plot the richness of detected groups as a function of redshift. 
In Figure~\ref{fig:lum_in_groups}, we have plotted the total luminosities of the same groups. Finally, in Figure~\ref{fig:veldisp_in_groups}, we give the velocity dispersion of the galaxies within these groups. As expected
the mean velocity dispersion does not vary significantly with distance, and has a mean value of  $\sim 95$~\kms.
The richness is approximately  constant up to $\sim 150$\Mpch{}, which is a design feature of the group finder. The minimal luminosity of the groups increases with distance, as we are losing the fainter objects at larger distances because the 2M++ catalogue is limited in apparent magnitude. The catalogue of group properties is given in Appendix~\ref{app:2mpp_group}. We have checked that the parameters of the fitted Schechter LF do not change significantly after the grouping of the galaxies.

\begin{figure}
  \includegraphics[width=\hsize]{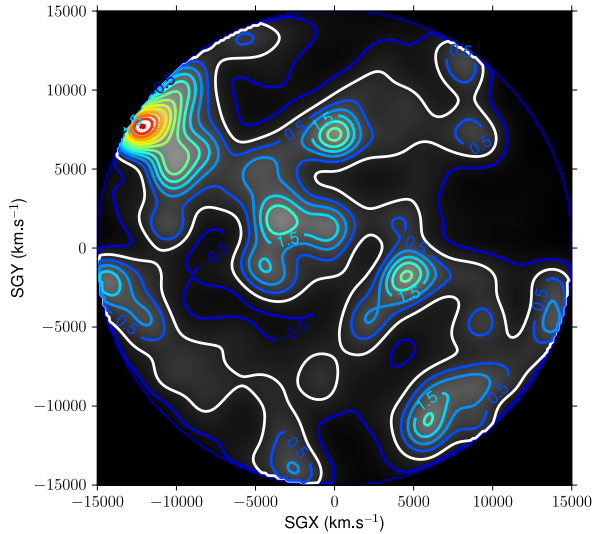}
  \caption{\label{fig:sg_density} The 2M++ number density field in Supergalactic plane.  The density field is smoothed with a Gaussian kernel of 1,000~\kms{} radius. Colour contours show the overdensity in units of the mean density and are separated by $0.5$.The mean density is highlighted by a white contour. }
\end{figure}

\begin{figure*}
  \includegraphics[width=\hsize]{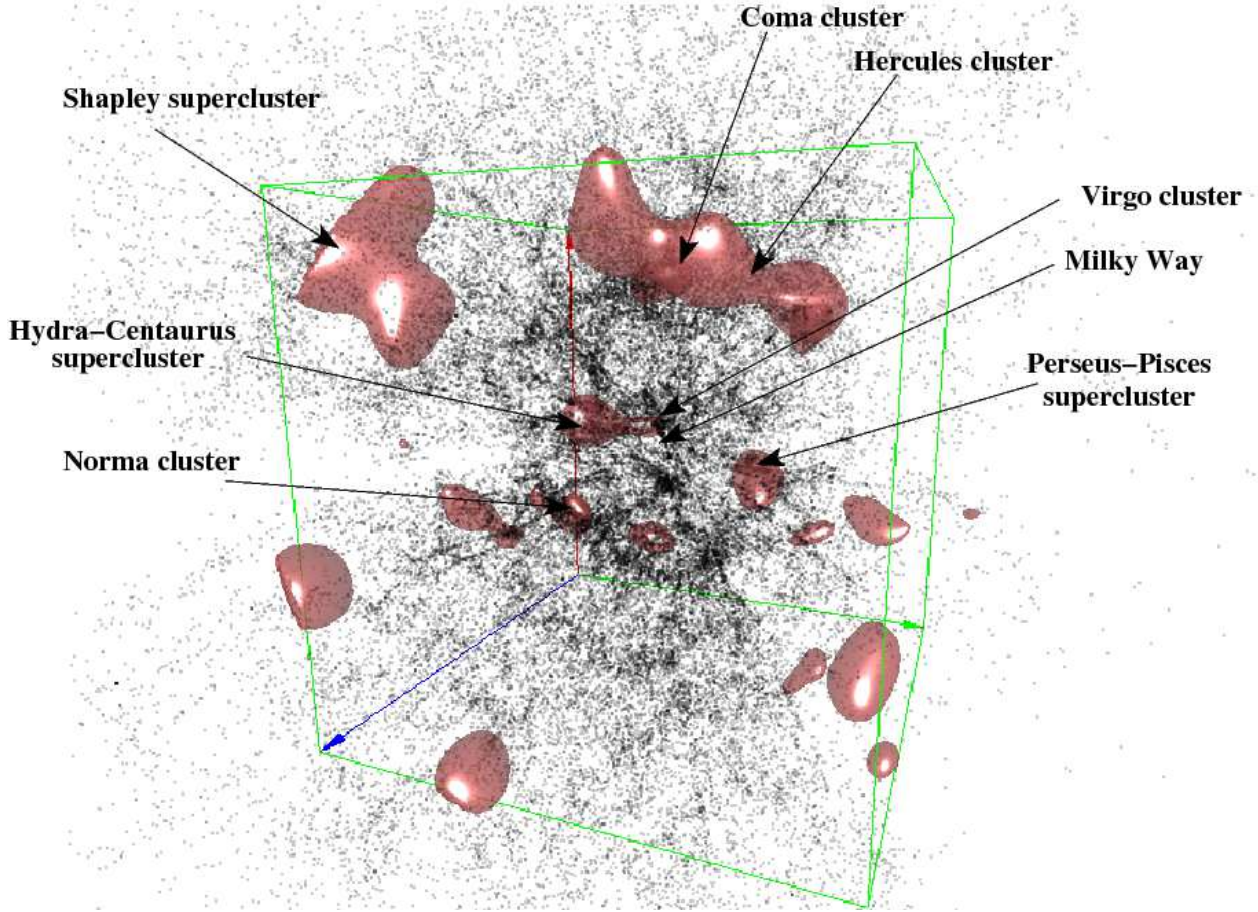}
  \caption{\label{fig:3d_density} The 2M++ galaxy distribution and density field in three dimensions. The cube frame is in Galactic coordinates. The Galactic plane cuts orthogonally through the middle of the back vertical red arrow. The length of a side of the cube is 200\Mpch{} and is centred on Milky Way.  We highlight the iso-surface of number fluctuation, smoothed with a Gaussian kernel of radius 1,000\kms{}, $\delta_\text{L}=2$ with a shiny dark red surface. The position of some major structures in the Local Universe are indicated by labelled arrows. We do not show isosurfaces beyond a distance of 150\Mpch{}, so Horologium-Reticulum is, for example, not present.}
\end{figure*}

\section{Density field}
\label{sec:density}

In this Section, we consider some properties of the peaks in the three-dimensional density field obtained from the distribution of galaxies in the 2M++ galaxy redshift catalogue. We assume that the number density and luminosity density of galaxies follow a Poisson distribution. As such, the mean smoothed density contrast $\rho(\mathbf{x})$ given the galaxy weights $w_i$ is
\begin{equation}
	\rho(\mathbf{s}) = \frac{1}{\bar{\rho}} \sum_{i=1}^{N_\text{galaxies}} W(\mathbf{s}-\mathbf{s}_i) w_i
\end{equation}
and the standard deviation
\begin{equation}
	\sigma^2_\rho(\mathbf{s}) = \frac{1}{\bar{\rho}^2} \sum_{i=1}^{N_\text{galaxies}} W(\mathbf{s}-\mathbf{s}_i)^2 w_i^2, \label{eq:error_poisson}
\end{equation}
with $\mathbf{s}$ the coordinate in redshift space, $W(\mathbf{x})$ the smoothing kernel considered. To compute the position of the peaks in this density field, we use an iterative spherical overdensity algorithm:
\begin{enumerate}
	\item we initialize the algorithm with an approximation $\mathbf{x}^0_\text{c}$ of the expected position of the cluster;
	\item we compute the barycenter $\mathbf{x}^{N+1}_\text{c}$ of the set of galaxies contained in a sphere centred on $\mathbf{x}^N_\text{c}$ and with radius $R_N$;
	\item we iterate (ii) until convergence, setting $R_{N+1}=R_N$;
	\item we reduce $R_{N+1}=0.80 R_N$. If $R_{N+1} > 1$\Mpch{}, then  we go back to step (ii), in the other case we terminate the algorithm.
\end{enumerate}
We define the position of the structure as the one given by the last step in the above algorithm. This position is used in the following sections to compute mean densities and infall velocities on clusters.

\begin{figure*}
	\includegraphics[width=\hsize]{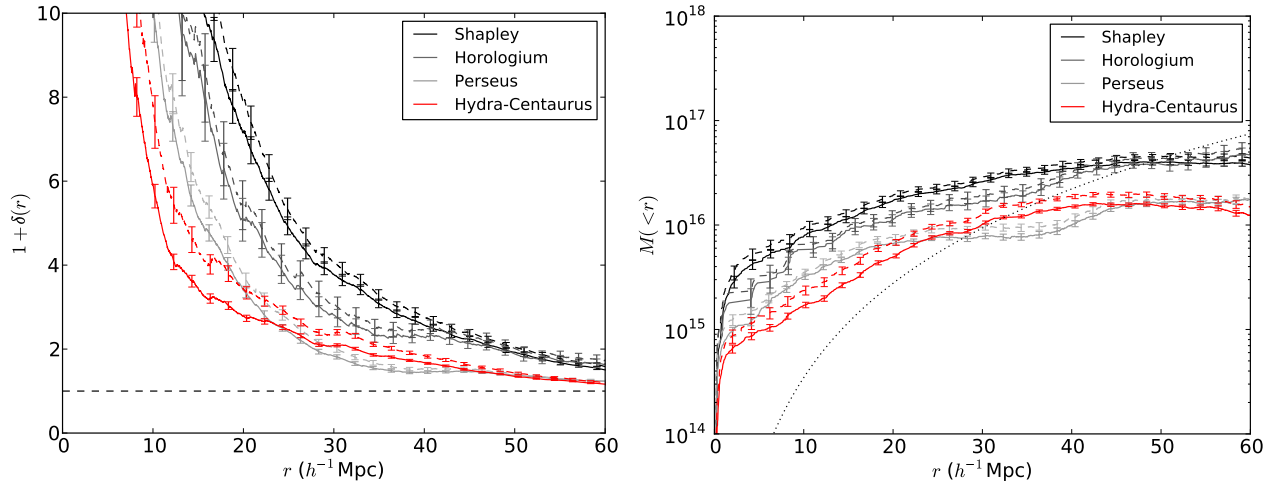}
	\caption{\label{fig:cluster_profiles} The cumulative average density profile and the excess mass as a function of radius from four major superclusters in the 2M++ redshift catalogue. In the two panels, we both show the profiles computed using the number weighed (solid lines) and the luminosity weighed (dashed lines) scheme. In the left panel, the horizontal black dashed line corresponds to the mean density. In the right panel, the dotted lines indicates the mass of a sphere of the given radius at the mean density. Note that since we are plotting excess mass, to obtain the total mass one must add this value. The black, dark grey, light grey and red lines correspond respectively to the Shapley concentration, the Horologium-Reticulum supercluster, the Perseus-Pisces supercluster and the Hydra-Centaurus supercluster. The error bars are estimated assuming that galaxies follow Poisson distribution for sampling the matter density field, as given by Eq.~\eqref{eq:error_poisson}.}
\end{figure*}

\subsection{Cosmography}
\label{sec:cosmography}

Fig.~\ref{fig:sg_density} shows the galaxy number density field of our catalogue in the Supergalactic Plane, smoothed to 10\Mpch{} with a Gaussian kernel. The  Shapley concentration (SC) in the upper-left corner, near $(\text{SGX},\text{SGY})\simeq(-10000,7000)$~\kms{}, is particularly prominent and is the largest density fluctuation in the 2M++ catalogue. The Shapley region is covered by the 6dF portion of the survey which extends to a depth  $\ktmpp \le 12.5$. Shapley is thus correctly sampled and is not a result of overcorrection of data limited to $\ktmpp \le 11.5$.  When smoothed with a Gaussian kernel of 10\Mpch{} radius, the Shapley concentration peaks at  $(l,b)=(312,30)$ and $d=152$\Mpch{} with a density $1+\delta_g=8.83\pm 0.46$, in galaxy number density contrast, and $1+\delta_\text{L}=9.51\pm 0.54$ in terms of luminosity density contrast.

The second most important structure in the Supergalactic plane of the 2M++ catalogue is the Perseus-Pisces (PP) supercluster. It is clearly seen in the Supergalactic plane in Fig.~\ref{fig:sg_density} at $(\text{SGX},\text{SGY})\simeq(5000,-1000)$~\kms{}. Its highest redshift space density, smoothed with Gaussian kernel of 10\Mpch{} radius, is about $1+\delta_g=4.46\pm 0.18$ in terms of number density contrast, $1+\delta_\text{L}=4.47\pm 0.20$ in terms of luminosity density contrast. The position of the peak corresponds to the Perseus cluster at $(l,b)=(150,-13)$ , which is quite near the ZoA, and a distance of 52~\Mpch{}.  It is quite possible that the filling of the ZoA by galaxies cloned from the Perseus itself amplifies the overdensity of this supercluster. 

The extended overdense structure in the central part of the Supergalactic plane, at about $(\text{SGX},\text{SGY})\simeq(-5000,0)$~\kms{}, is the Hydra-Centaurus-Virgo (HC) supercluster. At 10\Mpch{} smoothing scale, the highest peak, located at $(l,b)=(302,21)$, $d=38$\Mpch{}, coincides with the Centaurus cluster and has a height of $1+\delta_g = 3.02\pm 0.08$ in number density contrast and $1+\delta_\text{L}=3.40\pm 0.14$ is luminosity density contrast. 

Finally, Fig.~\ref{fig:3d_density} is a three-dimensional representation of the catalogue in Galactic coordinates, which means the Galactic plane goes through the middle of the vertical sides of the box, near the Norma cluster. We plot the 2M++ galaxies as points. Strong overdensities are highlighted by a transparent dark-red iso-surface of density fluctuation of luminosity $\delta_\text{L}=2$. This density has been smoothed at 10\Mpch{} with a Gaussian kernel from the corrected number distribution. The Shapley supercluster is located at the top-left corner of the cube. A number of overdensities in the right part of the cube arise from the high weights, as this region has a depth of only $\ktmpp \leq 11.5$.

\subsection{Supercluster masses}
\label{sec:masses}

We show in Fig.~\ref{fig:cluster_profiles} the mean overdensity and excess mass within a sphere of 50 \Mpch{} for four important superclusters in the 2M++ catalogue: the Shapley concentration, the Perseus-Pisces supercluster, the Horologium-Reticulum (HR) supercluster centred at $(l,b)=(265,-51)$ and a distance of 193 \Mpch{} and the Hydra-Centaurus supercluster. The profiles are centred on the position where the density peaks for each supercluster. 

For the four superclusters, we note that the profiles obtained through number weighing and luminosity weighing are nearly equivalent. The bumps in the mean density, shown in the left panels, are reproduced in both weighing schemes. This is particularly striking for the PP supercluster, even for scales as small as 10\Mpch{}. 
In all cases, the luminosity weighted contrast is slightly lower than the number weighted density contrast. In the following discussion, we adopt the luminosity-weighted number contrast.

The excess masses of all superclusters converge at radii of $\sim 50$ \Mpch{}. While Shapley is the most massive supercluster, we find that HR is very similar when measured on scales of 50 \Mpch{}.  Both have masses close to $10^{17}$ \hMsun.  The PP and HC superclusters are less massive, but, being considerably closer, these have more impact in the motion of the LG and nearby galaxies, as we discuss below.

Our estimate of Shapley's mass and density contrast is similar to that of \citet{ProQuiCar06} who measured a density contrast $\delta_n = 5.4\pm0.2$ in a truncated cone of 225 square degrees between 90 and 180 \Mpch{} with a volume equivalent to a sphere of effective radius 30.3 \Mpch{}.  In a sphere of this radius centred on Shapley, we find a luminosity density contrast of $\delta_{K,L} = 4.1\pm0.15$.

\citet{MunLoe08} calculated the mass of SC based on the overdensity of rich clusters and obtained a mass $3.3\pm0.3\ten{16}$ \hMsun{} within a sphere of 35 \Mpch{}. On the same scale, we obtain a mass of $4.87\pm0.18\ten{16}$ \hMsun, assuming $\Omega_\text{m}=0.3$ and $b_{K,L}=1$ for $K$-band luminosity. Using similar arguments, \citet{SheDia11} quote a mass of $1.8\ten{16}$ \hMsun{} within a slightly smaller radius of 31 \Mpch{}. On the same scale we find $4.00\pm0.17\ten{16}$ \hMsun.  These values could be brought into rough agreement if luminosity-weighted 2MASS galaxies are strongly biased, with $b = 2$ -- 3.  Such a strong biasing would, however, conflict with the measurement $b_{K,n} = 1.05\pm0.10 (\Omega_\text{m}/0.3)^{0.55}$ by \citet{PikeHudson05} but may be marginally consistent with the lower value $b_{K,n} = 1.56\pm0.16 (\Omega_\text{m}/0.3)^{0.55}$ found recently by \citet{DavNusMas10}.

A further caveat is that our density estimates are in redshift-space, and so are enhanced by a factor up to $b_s=1.2$ \citep{Kaiser87}\footnote{$b_s=\sqrt{1 + 2 f/3 + f^2/5}$, with $f=0.5$.} compared to the estimates of \citet{MunLoe08} and \citet{SheDia11}. In a future paper, we will reconstruct the density field in real-space and calibrate the biasing factor directly using peculiar velocity data, so a detailed comparison of overdensities awaits future work.

\citet{HSLB04} studied the overdensity of the SC as traced by IRAS-selected galaxies.   Within a 50 \Mpch{}-radius sphere they found that the overdensity of IRAS-selected galaxies is only 0.2. Here we find that the overdensity of 2MASS-selected galaxies on the same scale is $\sim 1$. Clearly, the relationship between IRAS and 2MASS-selected galaxies is not well-described by a relative linear bias, since a value of $\sim 5$ would be required in the Shapley supercluster, whereas the field requires a relative bias between 2MASS- and IRAS-selected galaxies of $\sim 1$ \citep{PikeHudson05}. 

\begin{table*}
  \begin{center}
  \begin{tabular}{lrrrrr}
    \hline
    Supercluster   & \multicolumn{3}{c}{Sphere centre} & $1+\delta_L$ & Mass \\
    & r & $l$ & $b$ \\
                     & \Mpch & $^{\circ}$ & $^{\circ}$ & & $10^{16}$ \hMsun\\
                    
    \hline
    Shapley         & 152 & 312 & 30 & $2.05\pm0.05$ & $8.9\pm0.2$\\
    Horologium-Reticulum      & 193 & 265 & $-51$ & $2.01\pm0.10$ & $8.7\pm0.5$ \\
    Perseus-Pisces  & 52 & 150 & $-13$ & $1.41\pm0.03$ & $6.1\pm0.1$ \\
    Hydra-Centaurus & 38 & 302 & 21 & $1.43\pm 0.03$ & $6.2\pm0.1$ \\
    \hline
  \end{tabular}
  \end{center}
  \caption{\label{tab:masses} Luminosity density contrast and estimated masses (assuming $\Omega_\text{m}=0.3$ and $b_{K,L} = 1$) of the four superclusters from the distribution of galaxy light within a sphere of radius 50 \Mpch{}.}
\end{table*}

\begin{figure}
	\includegraphics[width=\hsize]{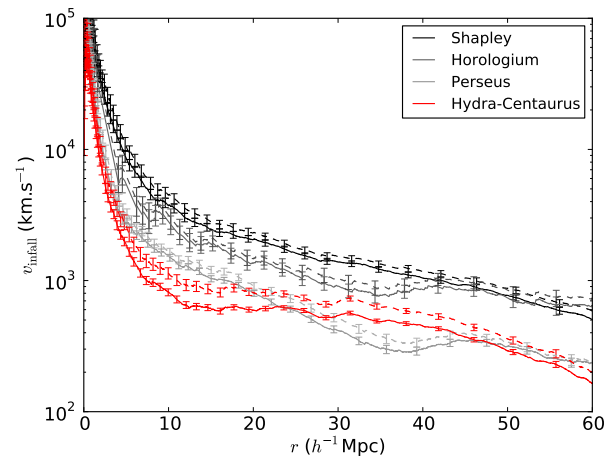}
	\caption{\label{fig:cluster_infall} The infall velocities as a function of distance for four major superclusters in the 2M++ redshift catalogue. Curves and error bars are as in Fig. \ref{fig:cluster_profiles}.}
\end{figure}

\subsection{Supercluster infall}
\label{sec:infall}

We now discuss the impact these structures have on large-scale flows in the nearby Universe. We have estimated the infall velocity onto each of these structures using linear theory:
\begin{equation}
	v_\text{infall} = \frac{1}{3} \beta H \bar{\delta}(R) R,
\end{equation}
with $H$ the Hubble constant, $\beta \equiv f/b$ where $f$ is the linear density perturbation growth rate and $b$ is a biasing parameter, and $\bar{\delta}(R)$ the mean density inside a sphere of radius $R$ and centred on the supercluster. 
For a \LCDM{} cosmology, $f \simeq \Omega_\text{m}^{5/9}$ \citep{BCHJ95}. We use $\beta=0.5$ \citep[as determined for 2MASS galaxies by][]{PikeHudson05} whenever we need to estimate a velocity. This value corresponds to $\Omega_\text{m}\simeq 0.30$ and $\Omega_\Lambda=1-\Omega_\text{m}$ with $b = 1$.

Fig.~\ref{fig:cluster_infall} shows the infall velocity profiles of
the four superclusters. Although we plot the linear theory infall down
to small radii ($R \la 10$\Mpch{}), we note that linear theory does
not apply in these regions and focus the discussion on distances $R
\ga 10$\Mpch{}. The infall velocities at 10\Mpch{} are all at least
2,000~\kms{}, with Shapley having the highest infall at nearly
4,000~\kms{}. At 50\Mpch{}, the Shapley and HR superclusters have an
infall of $\sim$800~\kms{}. The average overdensity of the Shapley
concentration within a sphere of 50\Mpch{} is
$1+\delta_\text{L}=2.05\pm 0.05$. Neglecting structures beyond 50
\Mpch{}, linear theory implies that the supercluster is responsible
for attracting the LG with a peculiar velocity of $90\pm 10$\kms{}. This
motion represents $\sim 15\%$ of the total velocity of the LG
with respect to the CMB rest frame. Although
the excess mass of HR is similar to Shapley, its effect on the LG's
motion is less than that of Shapley due to its greater distance: we
estimate 60 \kms{}. Added vectorially, the net peculiar velocity from
these two superclusters is approximately
110 \kms{} towards $(l,b) = (297,-1)$. This direction is within the
errors of the direction of the 407 \kms{} bulk flow found by \citet{WFH09}, but is
lower in amplitude.

Because they are closer to the LG, the HC and PP superclusters have a
greater impact. Approximating HC as a sphere, the infall at the LG's
distance of 38 \Mpch{} is $588\pm26$\kms. Whereas PP is denser, its
greater distance of 52 \Mpch{} puts it on the losing side of the
gravitational tug-of-war with HC: the infall of the LG towards PP is
only $313\pm24$ \kms{}. Thus the net motion is towards HC.

Note that it is likely that underdense regions also contribute a push. \citet{KocEbe06} have noted the deficit of rich clusters in the Northern sky, particularly in the distance range 130 to 180 \Mpch{}. Thus a full analysis of peculiar velocities requires integration over the entire density field, a topic we defer to a later paper.
 


\section{Summary}
\label{sec:conclusion}

We have compiled a new, nearly full-sky galaxy redshift catalogue, dubbed 2M++, based on the data from three redshift surveys: the 2MASS Redshift Survey ($K\le 11.5$), the Sloan Digital Sky Survey and the 6dF galaxy redshift survey. After having calculated corrected magnitudes and having calculated redshift completeness, we have determined LFs and weights that allow us to determine the redshift density field to a depth of 200 \Mpch{}. The most prominent structure within  200 \Mpch{} is the Shapley Concentration: its luminosity density within a sphere of radius 50\Mpch{} is 2.05 times the mean, and is thus responsible for approximately 90~\kms{} of the LG's motion with respect to the CMB rest frame. We have compared the density profile of four massive superclusters that are present in the 2M++ catalogue: the Shapley Concentration, the Perseus-Pisces supercluster and the Horologium-Reticulum supercluster and Hydra-Centaurus. The Shapley Concentration is clearly the most massive of the four, but HR is only slightly less massive.

This new, deep full-sky catalogue will be used in future work to study the peculiar velocity of the LG and other nearby galaxies. Our hope is that the distribution of density in the 2M++ volume will account for the high-amplitude bulk motions on scales of 100\Mpch{} \citep{WFH09,Lavaux10}.

\section*{Acknowledgements}

We thank Roya Mohayaee for stimulating discussions, and for helping to initiate this project at the Institut d'Astrophysique de Paris. We also thank the IAP for its hospitality.

This publication makes use of data products from the Two Micron All Sky Survey, which is a joint project of the University of Massachusetts and the Infrared Processing and Analysis Center/California Institute of Technology, funded by the National Aeronautics and Space Administration and the National Science
Foundation. 

We thank Lucas Macri and the 2MRS team for making their redshifts available in advance of publication.

We thank the 6dFGS team for the survey and for making their redshifts available at http://www.aao.gov.au/local/www/6df/

Funding for the SDSS and SDSS-II has been provided by the Alfred P. Sloan Foundation, the Participating Institutions, the National Science Foundation, the U.S. Department of Energy, the National Aeronautics and Space Administration, the Japanese Monbukagakusho, the Max Planck Society, and the Higher Education Funding Council for England. The SDSS Web Site is http://www.sdss.org/.

The SDSS is managed by the Astrophysical Research Consortium for the Participating Institutions. The Participating Institutions are the American Museum of Natural History, Astrophysical Institute Potsdam, University of Basel, University of Cambridge, Case Western Reserve University, University of Chicago, Drexel University, Fermilab, the Institute for Advanced Study, the Japan Participation Group, Johns Hopkins University, the Joint Institute for Nuclear Astrophysics, the Kavli Institute for Particle Astrophysics and Cosmology, the Korean Scientist Group, the Chinese Academy of Sciences (LAMOST), Los Alamos National Laboratory, the Max-Planck-Institute for Astronomy (MPIA), the Max-Planck-Institute for Astrophysics (MPA), New Mexico State University, Ohio State University, University of Pittsburgh, University of Portsmouth, Princeton University, the United States Naval Observatory, and the University of Washington.

This research has made use of the NASA/IPAC Extragalactic Database (NED) which is operated by the Jet Propulsion Laboratory, California Institute of Technology, under contract with the National Aeronautics and Space Administration.

GL acknowledges financial support from NSF Grant AST 07-08849.  MH acknowledges support from NSERC.  GL and MH acknowledge the financial support of the ``Programme visiteur de l'IAP'' and the French ANR via the OTARIE project.

Finally, we acknowledge the lifetime work of John Huchra, a pioneer of galaxy redshift surveys, as well as the driving force behind the 2MASS Redshift Survey that is such a critical part of the dataset compiled here. He will be missed.


\input{biblio}
\appendix

\section{The 2M++ data}

\makeatletter
\let\@makecaption=\SFB@maketablecaption
\makeatother

\begin{sidewaystable*}
    \centering
    \begin{tabular}{ccccccccccccccccc}
      \hline
      Name & $l$ & $b$ &  $\ktmpp$ & $V_\text{helio}$ & $V_\text{CMB}$ & $V_\text{err} $ & Group & $c_{11.5}$ & $c_{12.5}$ & ZOA & Cloned & $M_0$ & $M_1$ & $M_2$  & Bibcode\\ 
           & (deg.) & (deg.) & & (\kms) & (\kms) & (\kms) & & & & & & & \\
      (1)  & (2) & (3) & (4) & (5) & (6) & (7) & (8) & (9) & (10) & (11) & (12) & (13) & (14) & (15) & (16)  \\
      \hline
      \hline
\input{data1}
      \hline
    \end{tabular}
    \caption{\label{tab:2mpp} {\it The 2M++ catalogue} --
    Col. (1): the name of the galaxy as given in the 2MASS-XSC database. 
    Col. (2): Galactic longitude in degrees. 
    Col. (3): Galactic latitude in degrees. 
    Col. (4): Apparent magnitude in band $K_\text{S}$ as
    defined in Section~\ref{sec:magcor}. 
    Col. (5): Heliocentric total apparent velocity. 
    Col. (6): Total apparent velocity in CMB rest frame, using relation from \protect\cite{K93} and \protect\cite{Tully08}. 
    Col. (7): Total apparent velocity error (equal to zero if not measured). 
    Col. (8): Unique group identifier obtained from the algorithm of Section~\ref{sec:groups}. 
    Col. (9): Redshift incompleteness at magnitude $\ktmpp\le 11.5$. 
    Col. (10): Redshift incompleteness at magnitude $\ktmpp \le 12.5$. It may be
               empty in that case the catalogue is limited to $\ktmpp \le 11.5$ in 
               the portion of the sky holding the galaxy. 
    Col. (11): Flag  to indicate whether this is is a fake galaxy to fill the zone of
               avoidance following the algorithm of Section~\ref{sec:zoa}. 
    Col. (12): Flag to indicate if the redshift has been obtained by the cloning procedure of Section~\ref{sec:red_cloning}. 
    Col. (13): Flag to indicate whether this galaxy lies in the exclusive region covered by the 2MRS target mask (2Mx6S region).
    Col. (14): Flag to indicate whether this galaxy lies in the non-exclusion region covered by the SDSS.
    Col. (15): Same as (14) but for the 6dFGRS.
    Col. (16): Bibliographic code for the origin of the redshift information. The code is truncated in the above table but available in full in the electronic version of the catalog.
    }
\end{sidewaystable*}

\section{The 2M++ group catalog}
\label{app:2mpp_group}

\begin{sidewaystable*}
    \centering
    \begin{tabular}{cccccccc}
      \hline
      Id & $l$ & $b$ & $\ktmpp$ & $N_\mathrm{galaxies}$ & $V_\mathrm{helio}$ & $V_\mathrm{CMB}$ & $\sigma_V$ \\
         & (deg.) & (deg.) &  &                     &  (\kms)             & (\kms)           & (\kms) \\ 
      (1) &  (2)  &  (3)   & (4)& (5)               & (6)                 & (7)              & (8) \\
      \hline
      \hline
      \input{group1}
      \hline
    \end{tabular}
    \caption{\label{tab:2mpp_group} {\it The 2M++ group catalogue} --
      Col. (1): Group identifier in the catalogue. It corresponds to column 8 of Table~\ref{tab:2mpp}.
      Col. (2): Galactic longitude
      Col. (3): Galactic latitude
      Col. (4): Apparent magnitude in band $K_\text{S}$ as defined in Section~\ref{sec:magcor}. The magnitude is derived from the 2M++ galaxies. This is a magnitude uncorrected for incompleteness effect.
      Col. (5): Richness, uncorrected for incompleteness effect.
      Col. (6): Heliocentric total apparent velocity.
      Col. (7): Total apparent velocity in CMB rest frame, using relation from \protect\cite{K93} and \protect\cite{Tully08}.
      Col. (8): Velocity dispersion in the group.
    }
\end{sidewaystable*}

\end{document}

%% file: data1.tex
07345116-6917029 & 281.00 & -21.54 & 7.90 & 1367 & 1486 & 69 & 4996 & 1.0 & 1.0 & 0 & 0 & 0 & 0 & 1 & 20096dF..... \\ 
21100305-5448123 & 342.30 & -41.63 & 12.31 & 18717 & 18571 & 0 &  & 1.0 & 0.9 & 0 & 0 & 0 & 0 & 1 & 20096dF..... \\ 
20353522-4422308 & 356.19 & -36.74 & 11.44 & 7066 & 6893 & 198 & 4388 & 1.0 & 1.0 & 0 & 0 & 0 & 0 & 1 & 20096dF..... \\ 
13271270-2451409 & 313.15 & 37.30 & 12.04 & 12132 & 12429 & 0 &  & 1.0 & 1.0 & 0 & 0 & 0 & 0 & 1 & 20096dF..... \\ 
21112498-0849375 & 41.49 & -35.05 & 12.37 & 8296 & 7988 & 0 & 4177 & 1.0 & 0.9 & 0 & 0 & 0 & 0 & 1 & 20096dF..... \\ 
02581778-0449064 & 182.17 & -52.44 & 10.50 & 9235 & 9036 & 0 & 3733 & 1.0 & 0.9 & 0 & 0 & 0 & 0 & 1 & 20096dF..... \\ 
14303940+0716300 & 357.34 & 59.22 & 9.97 & 1370 & 1601 & 10 &  & 1.0 & 0.8 & 0 & 0 & 0 & 1 & 0 & 1998AJ...... \\ 
00362801+1226414 & 117.21 & -50.26 & 11.44 & 10339 & 9993 & 11 &  & 0.8 & 0.0 & 0 & 0 & 1 & 0 & 0 & 20112MRS.... \\ 
18490084+4739293 & 77.11 & 20.08 & 10.40 & 4671 & 4536 & 10 &  & 0.8 & 0.0 & 0 & 0 & 1 & 0 & 0 & 1991RC3.9... \\ 
23054906-8545110 & 305.13 & -30.91 & 12.45 & 19770 & 19788 & 270 &  & 1.0 & 0.7 & 0 & 1 & 0 & 0 & 1 &     none \\ 
07511205-8540159 & 298.28 & -25.94 & 12.44 & 25990 & 26053 & 270 &  & 1.0 & 0.6 & 0 & 1 & 0 & 0 & 1 &     none \\ 
03355460-8537067 & 299.59 & -30.39 & 11.90 & 12702 & 12736 & 270 & 3775 & 1.0 & 0.7 & 0 & 1 & 0 & 0 & 1 &     none \\ 
08423963-8430223 & 297.59 & -24.47 & 11.70 & 12284 & 12357 & 270 & 4039 & 1.0 & 1.0 & 0 & 1 & 0 & 0 & 1 &     none \\ 
08431792-8429053 & 297.58 & -24.44 & 12.35 & 12284 & 12357 & 270 & 4039 & 1.0 & 1.0 & 0 & 1 & 0 & 0 & 1 &     none \\ 
08424060-8427453 & 297.55 & -24.44 & 11.96 & 12284 & 12357 & 270 & 4039 & 1.0 & 1.0 & 0 & 1 & 0 & 0 & 1 &     none \\ 
ZOA0000000 & 330.07 & -0.41 & 10.21 & 5238 & 5387 & 0 &  & 0.9 & 0.7 & 1 & 0 & 0 & 0 & 1 &      zoa \\ 
ZOA0000001 & 330.49 & -0.33 & 11.31 & 8841 & 8989 & 0 &  & 0.9 & 0.6 & 1 & 0 & 0 & 0 & 1 &      zoa \\ 
ZOA0000002 & 330.01 & -1.18 & 11.96 & 11830 & 11983 & 0 &  & 1.0 & 0.9 & 1 & 0 & 0 & 0 & 1 &      zoa \\ 
ZOA0000003 & 331.29 & -1.10 & 11.09 & 3158 & 3306 & 70 &  & 0.9 & 0.8 & 1 & 0 & 0 & 0 & 1 &      zoa \\ 
07243410-8543223 & 298.20 & -26.43 & 11.51 & 5301 & 5361 & 69 & 4638 & 1.0 & 0.7 & 0 & 0 & 0 & 0 & 1 & 20096dF..... \\ 
03403012-8540119 & 299.56 & -30.29 & 12.41 & 12714 & 12749 & 66 & 3775 & 1.0 & 0.6 & 0 & 0 & 0 & 0 & 1 & 20096dF..... \\ 
07400785-8539307 & 298.20 & -26.13 & 11.09 & 5184 & 5246 & 11 & 4638 & 1.0 & 0.7 & 0 & 0 & 0 & 0 & 1 & 2008ApJ..... \\ 
03355460-8537067 & 299.59 & -30.39 & 11.90 & 12702 & 12736 & 270 & 3775 & 1.0 & 0.7 & 0 & 1 & 0 & 0 & 1 &     none \\ 
07420104-8525161 & 297.96 & -26.04 & 10.31 & 5150 & 5213 & 14 & 4638 & 1.0 & 0.7 & 0 & 0 & 0 & 0 & 1 & 20096dF..... \\ 
02090195-8520255 & 301.12 & -31.51 & 11.52 & 12675 & 12699 & 11 & 3776 & 1.0 & 0.9 & 0 & 0 & 0 & 0 & 1 & 20112MRS.... \\ 

%% file: group1.tex
1 & 281.26 & 73.47 & 6.60 & 65 & 1223 & 1556 & 599 \\
1000 & 182.41 & -13.06 & 10.65 & 4 & 3 & -25 & 97 \\
1001 & 91.75 & 51.01 & 6.15 & 4 & 691 & 752 & 51 \\
1002 & 137.70 & 12.33 & 5.39 & 9 & 1139 & 1055 & 118 \\
1003 & 316.23 & -10.88 & 10.42 & 3 & -80 & 17 & 21 \\
1004 & 123.60 & 74.51 & 3.80 & 9 & -249 & -33 & 156 \\
1005 & 184.67 & 83.05 & 4.96 & 24 & 675 & 956 & 312 \\
1006 & 144.05 & 66.22 & 3.59 & 69 & 845 & 1049 & 227 \\
1007 & 171.51 & 32.81 & 8.58 & 4 & 506 & 646 & 46 \\
1008 & 108.51 & 58.06 & 5.83 & 6 & 186 & 302 & 109 \\
1009 & 319.07 & -12.21 & 4.75 & 4 & -146 & -65 & 89 \\
1010 & 144.70 & 36.20 & 3.55 & 5 & 92 & 157 & 82 \\
1011 & 33.58 & 14.01 & 8.21 & 4 & 1865 & 1780 & 101 \\
1012 & 41.38 & 14.94 & 8.00 & 3 & 2291 & 2188 & 12 \\
1013 & 134.90 & 32.70 & 6.94 & 4 & 1330 & 1348 & 51 \\
1014 & 103.67 & 33.03 & 8.10 & 4 & 1174 & 1129 & 63 \\
1015 & 41.06 & 12.68 & 9.01 & 3 & 2741 & 2626 & 93 \\
1016 & 45.04 & 17.71 & 8.24 & 3 & 2261 & 2162 & 60 \\
1017 & 150.98 & 5.93 & 8.69 & 4 & 5096 & 5028 & 112 \\
1018 & 147.84 & 7.90 & 8.10 & 6 & 4807 & 4736 & 237 \\
1019 & 129.32 & 8.92 & 7.37 & 5 & 3276 & 3146 & 203 \\
1020 & 103.21 & 12.39 & 7.94 & 5 & 2692 & 2523 & 131 \\
1021 & 50.91 & 6.89 & 8.34 & 4 & 4831 & 4659 & 193 \\
1022 & 69.13 & 8.13 & 8.96 & 4 & 4556 & 4359 & 81 \\
1023 & 75.62 & 6.03 & 8.22 & 4 & 4711 & 4497 & 94 \\
1024 & 269.12 & 5.57 & 9.91 & 3 & 5052 & 5324 & 75 \\
1025 & 264.14 & 7.22 & 8.89 & 4 & 4685 & 4965 & 234 \\
1026 & 275.43 & 8.94 & 9.48 & 3 & 4008 & 4291 & 88 \\
1027 & 264.36 & 8.40 & 8.29 & 7 & 4796 & 5081 & 121 \\
1028 & 295.46 & 8.81 & 8.36 & 4 & 4449 & 4701 & 113 \\